\documentstyle[12pt,epsf]{article}
\begin{document}
 
\newcommand{\be}{\begin{eqnarray}}
\newcommand{\ee}{\end{eqnarray}}
\begin{flushright}
LBNL-42004
\end{flushright}
\vspace{3mm} 
\begin{center}
 
{\Large
{Leading Charm in Hadron-Nucleus Interactions in the Intrinsic Charm Model
\footnote{
This work was supported in part by the Director, Office of Energy
Research, Division of Nuclear Physics of the Office of High Energy
and Nuclear Physics of the U. S. Department of Energy under Contract
Number DE-AC03-76SF00098.}
}} \\[8ex]

T. Gutierrez$^a$ and R. Vogt$^{a,b}$\\[2ex]
 $^a$Physics Department\\
 University of California at Davis\\
 Davis, California\quad 95616  \\
 and\\
 $^b$Nuclear Science Division\\
 Lawrence Berkeley National Laboratory\\
 Berkeley, California\quad 94720 \\[6ex]

{\bf ABSTRACT}
\end{center}
\begin{quote}
Leading charm hadrons produced in hadron-nucleus interactions cannot be 
adequately described within the parton fusion model.  Recent results on charm
baryon production in
$\Sigma^- A$ interactions at 330 GeV with the WA89 detector 
disagree with fusion predictions.  Intrinsic heavy
quark pairs in the $\Sigma^-(dds)$ wavefunction 
provide a simple mechanism for 
producing fast charm hadrons.  We calculate leading charm baryon production 
from $\Sigma^-$, $\pi^-$ and $p$ projectiles in a 
two component model combining parton fusion with intrinsic charm.  Final state
 $D^{-}$, 
$\Sigma_{c}^{0}$, $\Xi_{c}^{+}$, and $\Lambda_{c}^{+}$ $d\sigma/dx_{F}$ 
distributions and $D^-/D^+$, $D_s^-/D_s^+$ and $\Lambda_c^+/\overline
\Lambda^+_c$ asymmetries are compared to WA89 data.  Predictions are made for 
650 GeV $\Sigma^- A$ and $\pi^- A$ interactions in the SELEX detector at 
Fermilab and for 800 GeV $pA$ interactions.
\end{quote}
\vspace{2mm}
\begin{center}
PACS numbers: 12.38.Bx, 13.75.Ev, 14.20.Lg, 14.40.Lb
\end{center}
\newpage

\section{Introduction}

One of the most striking features of charm hadroproduction is the
leading particle effect:  the strong correlation between the quantum
numbers of the projectile and the
final-state charm hadron. For example, more $D^-$  than $D^+$ are
produced at large $x_F$ in $\pi^- A \rightarrow D^\pm X$ interactions
\cite{E791,Agb,Bar,WA82,E7692}. Such correlations are remarkable because they
explicitly contradict the perturbative QCD factorization theorem \cite{fact}
which predicts that heavy quarks hadronize through jet
fragmentation functions independent of the initial state.

While leading charm effects are well established for $D$ mesons, observations
of charm baryons are more rare \cite{E791lam}.  Two
experiments with $\Sigma^-(dds)$ beams promise to clarify the situation in the 
baryon sector.  The hyperon beam, with a strange valence quark, presents a 
unique opportunity to study the flavor dependence of leading charm
hadroproduction since both charm and charm-strange baryons can be leading.
The first, WA89, which directs a 330 GeV hyperon beam on carbon and
copper targets, has reported the $x_F$ distributions of $D^- (d\overline c)$,
$\Sigma_c^0 (ddc)$, $\Xi_c^+(usc)$, and $\Lambda_c^+(udc)$  \cite{wa89}
as well as the $D^-/D^+$, $D_s^-/D_s^+$ and
$\Lambda_c/\overline \Lambda_c$ production asymmetries \cite{wa89asym}. 
The second, SELEX \cite{SELEX}, has a large
acceptance for forward charm production, enhancing the charm baryon yield.
Their 650 GeV beam, approximately half $\pi^- (\overline u d)$ 
and $\Sigma^-$, promises to
improve current samples from both beams by up to an order of magnitude.
They also plan to study the $A$ dependence of leading charm.

In previous work \cite{VB,VBlam}, a QCD mechanism which
produces leading charm at large $x_F$ was introduced.  An important 
feature of the model is coalescence, the
process through which a charm quark hadronizes by
combining with quarks of similar rapidities, such as projectile spectator
valence quarks.  In a gauge theory
the strongest attraction is expected to occur when the spectators and the
produced quarks have equal velocities \cite{BGS}.  Thus the
coalescence probability should be largest at small relative rapidity
and rather low transverse momentum where the invariant mass  of
the $\overline Q q $ system is minimized, enhancing the binding amplitude.
 
This coalescence occurs in the initial state where the projectile 
wavefunctions of {\it e.g.}\ the $\pi^-$, $p$ and $\Sigma^-$
can fluctuate into Fock configurations containing a $c \overline
c$ pair such as $| \overline u d c \overline c
\rangle$, $|uu d c \overline c \rangle$ or $|dds c \overline c
\rangle$ respectively.  In these states, two or more gluons are
attached to the charm quarks, reducing the amplitude by ${\cal O}(\alpha_s^2)$
relative to parton fusion \cite{VB}.  The longest-lived fluctuations in
states with invariant mass $M$ have a lifetime of ${\cal O}( 2 P_{\rm
lab}/M^2)$ in the target rest frame where $P_{\rm lab}$ is the 
projectile momentum. Since
the comoving charm and valence quarks have the same rapidity in these states,
the heavy quarks carry a large fraction of the projectile momentum and can thus
readily combine to produce charm hadrons with large
longitudinal momentum. Such a mechanism can then dominate the
hadroproduction rate at large $x_F$. This is the underlying
assumption of the intrinsic charm model \cite{intc} 
in which the wavefunction fluctuations are initially
far off shell. However, they materialize as charm hadrons when light spectator
quarks in the projectile  Fock state interacts with the target
\cite{BHMT}. Since such interactions are strong, charm
production will occur primarily on the front face of the nucleus in
the case of a nuclear target.  Thus the intrinsic charm mechanism has a
stronger $A$ dependence than charm production by leading-twist fusion.
 
In this work,  we concentrate on the charm hadrons 
studied by WA89 and SELEX in order to further examine the
relationship between fragmentation and coalescence mechanisms. The
calculations are made within a two-component model: leading-twist
fusion and intrinsic charm \cite{VB,VBlam,VBH1,VBH2}. 
 
Leading particle correlations are also an integral part  of the
Monte Carlo program PYTHIA \cite{PYT} based on the Lund string
fragmentation model. In this model it is assumed that the heavy
quarks are produced in the initial state with relatively small
longitudinal momentum fractions by the leading twist fusion
processes. In order to produce a strong leading particle effect at
large $x_F$, the string has to accelerate the heavy quark as it
fragments into the final-state heavy hadron.  Such a mechanism
demands that
large changes of the heavy quark momentum take place in the final
state.  Other models of leading charm production by recombination in the
final-state have been suggested \cite{hwa,bedny}.  However, in this work we
will only compare our results with the commonly used PYTHIA Monte Carlo.

In this paper, we first discuss the conventional mechanism for charm production
at leading twist, parton fusion, and how the hyperon beam is taken into account
in the model.  Section 3 reviews the intrinsic charm model and describes the
extension of the model used in this work.  In section 4, we compare our results
on $\Sigma^- A$ interactions with the WA89 data and make predictions for SELEX
with $\Sigma^-$ and $\pi^-$ beams as well as $pA$ interactions.  Finally, we
summarize our results.

\section{Leading-Twist Charm Production}
 
In this section we briefly review the conventional leading-twist
model for the production of charm hadrons in $\Sigma^- p$, $pp$ and
$\pi^- p$ interactions.  In leading-twist QCD,
heavy quarks are produced by the fusion subprocesses $g g
\rightarrow Q \overline Q$ and $q \overline q \rightarrow Q
\overline Q$.   The factorization theorem \cite{fact} predicts that
fragmentation is independent of the quantum numbers of both the
projectile and target. We will also show the
corresponding distributions of charm hadrons predicted by the
PYTHIA model \cite{PYT}.
 
Our calculations are at lowest order in $\alpha_s$.  A constant
factor $K \sim 2-3$ is included in the fusion cross section since
the next-to-leading order $x_F$ distribution is larger than the
leading order distribution by an approximately constant factor
\cite{RV2}. An additional factor of two is included to obtain the single charm
distribution, twice the $c \overline c$ cross section.
Note that neither leading order production nor the next-to-leading
order corrections can produce flavor correlations \cite{Frix} such as those
observed in leading charm production.
 
The charm hadron $x_F$ distribution, where $x_F = (2m_T/\sqrt{s})
\sinh y$, has the factorized form \cite{VBH2}
\be
\frac{d\sigma}{dx_F} = \frac{\sqrt{s}}{2} x_a x_b \int H_{AB}(x_a,x_b)
\frac{1}{E_1}\ \frac{D_{H/c}(z_3)}{z_3}\ dz_3\, dy_2\, dp_T^2 \, \, ,
\label{ltfus}
\ee
where $a$ and $b$ are the initial partons, 1 and 2 are the produced charm
quarks with $m_c = 1.5$ GeV, and 3 and 4 are the final-state charm hadrons.
The convolution of the subprocess cross sections for $q \overline q$
annihilation and gluon fusion with the parton densities is included
in $H_{AB} (x_a, x_b)$,
\be
H_{AB}(x_a,x_b) =  \sum_q [f_q^A(x_a)  f_{\overline q}^B(x_b) +
f_{\overline q}^A(x_a) f_q^B(x_b)] \frac{d \widehat{\sigma}_{q
\overline q}}{d \hat{t}} + f_g^A(x_a) f_g^B(x_b) \frac{d
\widehat{\sigma}_{gg}}{d \hat{t}} \, \, ,
\label{hab}
\ee
where $A$ and $B$ are the interacting hadrons and the scale dependence of the
parton densities has been suppressed.
The subprocess cross sections can be found in Ref.~\cite{Ellis}.  Since we 
study $c \overline c$ production with several different projectiles,
we specify the general $q \overline q$ convolution for three light flavors:
\be
\lefteqn{ \sum_q [f_q^A(x_a)  f_{\overline q}^B(x_b) +
f_{\overline q}^A(x_a) f_q^B(x_b)] = u^A(x_a) \overline u^B(x_b) +
\overline u^A(x_a) u^B(x_b)} \nonumber \\ 
&  & + d^A(x_a)  \overline d^B(x_b) + \overline d^A(x_a) d^B(x_b) +
s^A(x_a)  \overline s^B(x_b) + \overline s^A(x_a) s^B(x_b)
\label{convol}
\ee
Parton distributions of the hyperon are not available.
However, using baryon number and momentum
sum rules, a set of parton distributions for the 
$\Sigma^{-}$ can be
inferred from the proton distributions:
\begin{eqnarray}
\int_0^1 u^p_v (x) \, dx & = & \int_0^1 d^{\Sigma^-}_v (x) \, dx = 2 \\
\int_0^1 d^p_v (x) \, dx & = & \int_0^1 s^{\Sigma^-}_v (x) \, dx = 1 \, \, .
\label{partsum}
\end{eqnarray}
We also identify $s^p(x) = u^{\Sigma^-}(x)$.  Similar relations
can be made for the sea quarks.  The gluon distributions are thus assumed to
be the same in
the $\Sigma^-$ and the proton.
Both the GRV LO 94 \cite{GRV94}
and MRS D$^{-'}$ \cite{MRS}
parton distribution functions with $\bar{u}\neq\bar{d}\neq\bar{s}$ were used. 
Other, older distributions with a
symmetric sea, $\bar{u}=\bar{d}=\bar{s}$, produce identical results for
$\Sigma^- p$ and $pp$ interactions.  
 
The fragmentation functions, $D_{H/c}(z)$, describe the
hadronization of the charm quark where $z = p_H/p_c$ is the
fraction of the charm quark momentum carried by the charm
hadron, assumed to be collinear to the charm quark. According to the
factorization theorem, fragmentation is independent of the initial state
and thus cannot produce flavor correlations, 
precluding a leading charm effect. This
uncorrelated fragmentation will be modeled by two extremes: a delta
function, $\delta(z -1)$, and the Peterson function \cite{Pete},
extracted from $e^+e^-$ data. 
The Peterson function, derived from a non-hadronic initial state,  predicts a
softer $x_F$ distribution than observed in hadroproduction, even at
moderate $x_F$ \cite{VBH2} since the fragmentation decelerates the
charm quark.
The parameters of the Peterson function we use here are taken from
$e^+ e^-$ studies of $D$ production \cite{Chirn}. 
Typically fits to charm baryon fragmentation functions suggest increased
deceleration of the charm quark in final-state baryons relative to mesons. 
On the other hand, the delta-function assumes that the charm quark coalesces
with a low-$x$ spectator sea quark or a low momentum secondary quark
with little or no momentum loss \cite{VBH2}. This
assumption is more consistent with low $p_T$ charm hadroproduction data
\cite{AgB1,AgB2,Bar2} than Peterson fragmentation.
 
\begin{figure}[htb]
\setlength{\epsfxsize=0.95\textwidth}
\setlength{\epsfysize=0.5\textheight}
\centerline{\epsffile{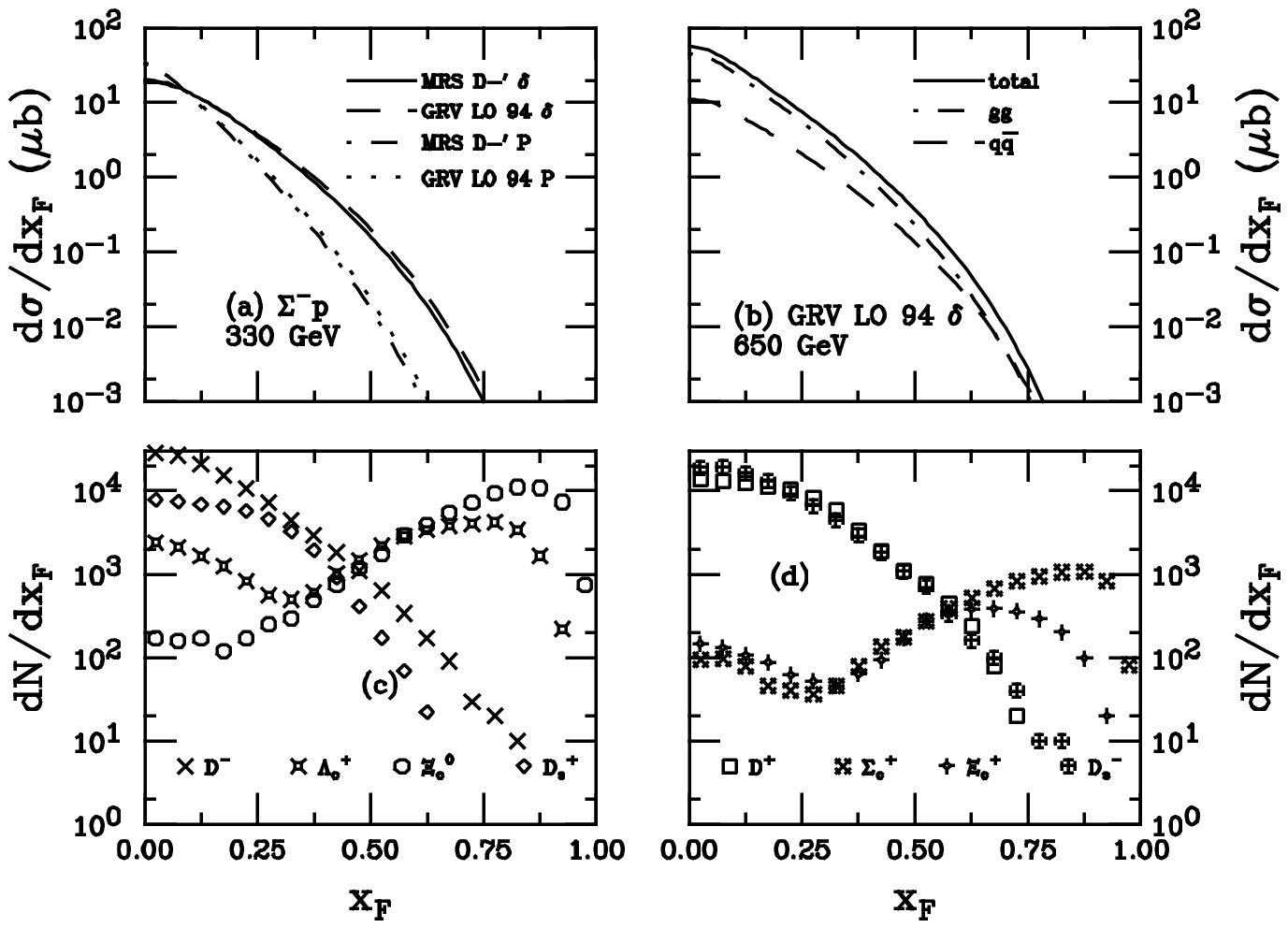}}
\caption[]{ Charm production by leading-twist fusion in $\Sigma^- p$
interactions.  (a) Two parton distribution functions with two different
fragmentation functions are shown at 330 GeV.  The curves show calculations 
with the MRS D$-^\prime$
parton distributions with delta function fragmentation (solid) and the Peterson
function (dot-dashed) and with the GRV LO 94 parton distributions with delta
function fragmentation (dashed) and the Peterson function (dotted).  In (b)
calculations with the GRV LO 94 parton distributions with delta function
fragmentation are given at 650 GeV for the $q \overline q$ component (dashed),
$gg$ component (dot-dashed) and the total production cross section (solid).
Charm hadron production in PYTHIA 6.115 at 330 GeV is shown in (c) and (d) 
with the distributions labeled as indicated.}
\label{sigfus}
\end{figure}

In Fig.~\ref{sigfus}(a) we show the  inclusive $x_F$ distributions
calculated for both types of fragmentation in $\Sigma^- p$ interactions at 
330 GeV.
Both sets of parton distributions are also shown.
Very little difference in either total cross section or shape
of the $x_F$ distributions can be discerned between the two sets of parton
distributions.  The delta function results in harder
distributions than those predicted by Peterson fragmentation for
$x_F > 0.2$. However, as shown in \cite{wa89}, even with this hard 
fragmentation  the fusion model cannot account
for the shape of leading charm baryon distributions.  Figure \ref{sigfus}(b) 
shows the 
relative contributions from $gg$ fusion and $q \overline q$ annihilation
to the total cross section at 650 GeV, the energy of the SELEX beam, using
the GRV LO 94 parton densities.  Gluon fusion clearly dominates the production
until $x_F \approx 0.6$.
We have checked that this is also true at the lower energy of the WA89 
experiment, 330 GeV.

\begin{figure}[htb]
\setlength{\epsfxsize=0.95\textwidth}
\setlength{\epsfysize=0.5\textheight}
\centerline{\epsffile{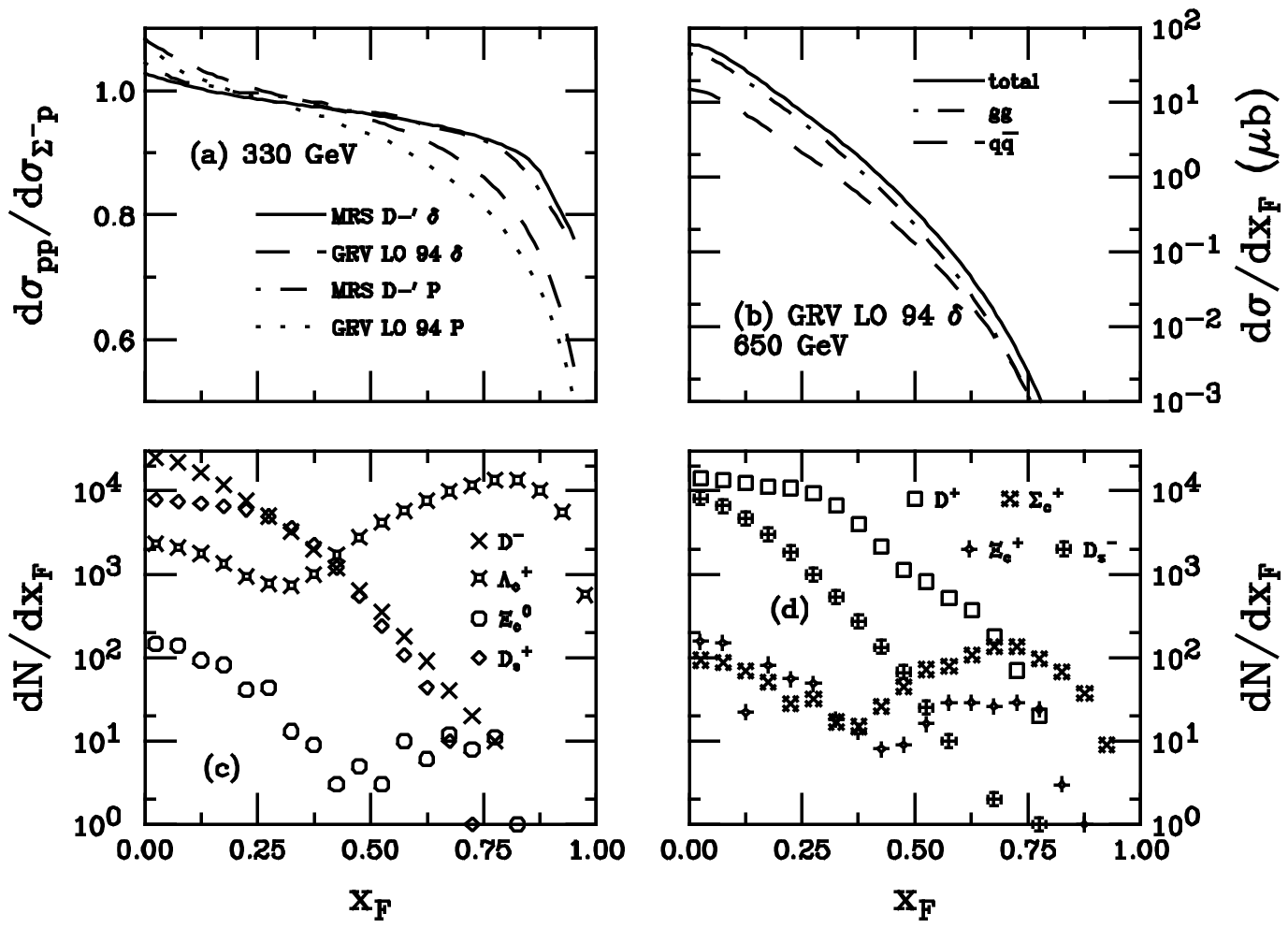}}
\caption[]{ Charm production by leading-twist fusion in $pp$
interactions.  (a) The cross section ratios $\sigma_{pp}/\sigma_{\Sigma^- p}$ 
are given for two parton distribution functions with two different
fragmentation functions at 330 GeV.  The curves show calculations 
with the MRS D$-^\prime$
parton distributions with delta function fragmentation (solid) and the Peterson
function (dot-dashed) and with the GRV LO 94 parton distributions with delta
function fragmentation (dashed) and the Peterson function (dotted).  In (b)
calculations with the GRV LO 94 parton distributions with delta function
fragmentation are given at 650 GeV for the $q \overline q$ component (dashed),
$gg$ component (dot-dashed) and the total production cross section (solid).
Charm hadron production in PYTHIA 6.115 at 330 GeV is shown in (c) and (d) 
with the distributions labeled as indicated.}
\label{profus}
\end{figure}

\begin{figure}[htb]
\setlength{\epsfxsize=0.95\textwidth}
\setlength{\epsfysize=0.5\textheight}
\centerline{\epsffile{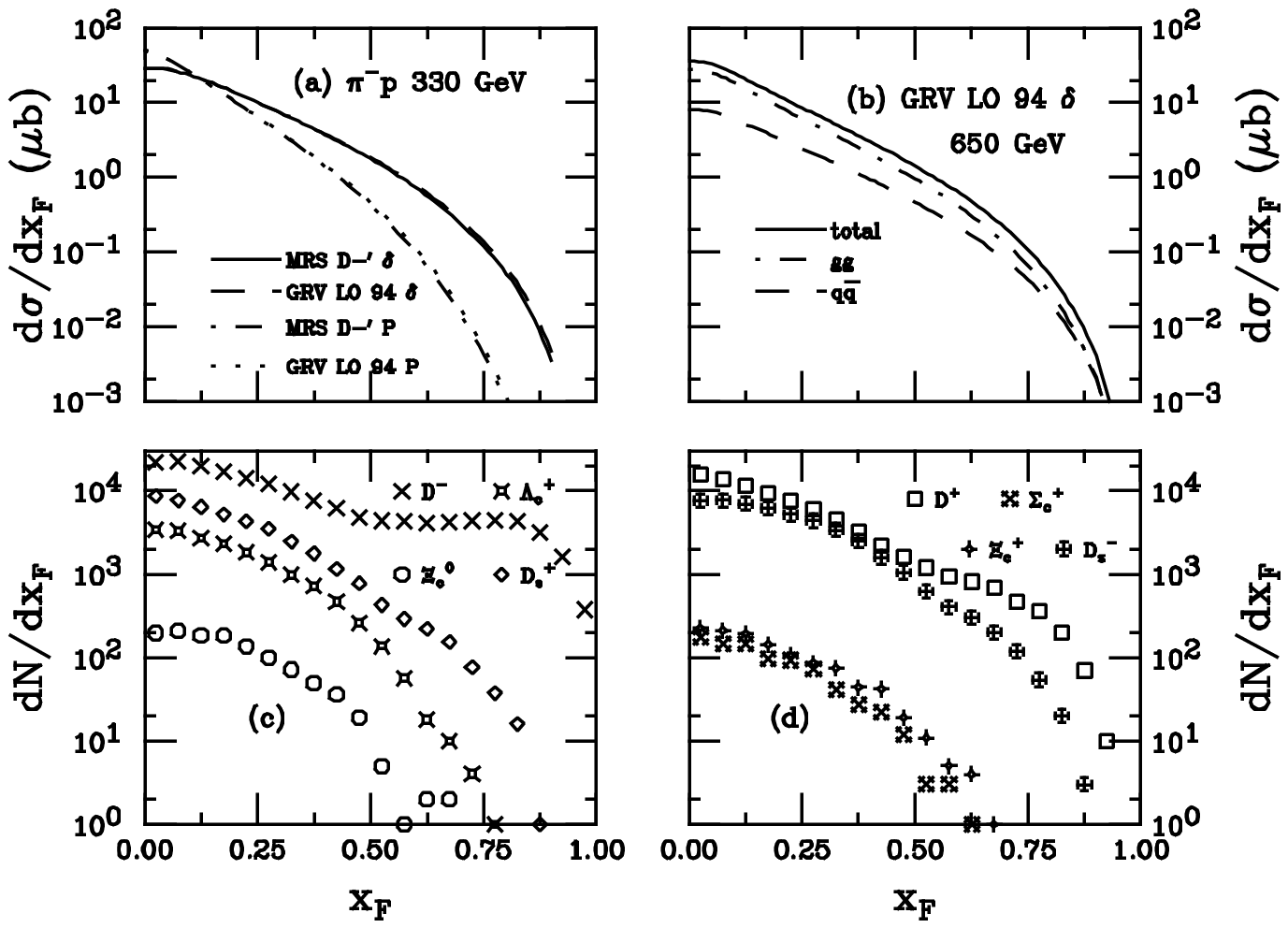}}
\caption[]{ Charm production by leading-twist fusion in $\pi^- p$
interactions.  (a) Two parton distribution functions with two different
fragmentation functions are shown at 330 GeV.  The curves show calculations 
with the MRS D$-^\prime$
parton distributions with delta function fragmentation (solid) and the Peterson
function (dot-dashed) and with the GRV LO 94 parton distributions with delta
function fragmentation (dashed) and the Peterson function (dotted).  In (b)
calculations with the GRV LO 94 parton distributions using delta function
fragmentation are given at 650 GeV for the $q \overline q$ component (dashed),
$gg$ component (dot-dashed) and the total production cross section (solid).
Charm hadron production in PYTHIA 6.115 at 330 GeV is shown in (c) and (d) 
with the distributions labeled as indicated.}
\label{pifus}
\end{figure}

We compare the
$\Sigma^- p$ distributions with those from $pp$ and $\pi^- p$ interactions at
the same energies with our two choices of parton distributions in 
Figs.~\ref{profus} and \ref{pifus} respectively.  
Since the differences between the $\Sigma^- p$
and $pp$ $x_F$ distributions are rather small due to the dominance of gluon
fusion, in Fig.~\ref{profus}(a) we show the ratio $\Sigma^- p/pp$ at 330 GeV.
The differences between $\Sigma^- p$ and $pp$ production are somewhat larger
for the GRV LO 94 distributions than the MRS D$-^\prime$, due to the relative
assumptions of $\overline u/ \overline d$.  The GRV LO 94 set is based on more
recent data than the MRS D$-^\prime$ and should thus more accurately reflect
the sea quark abundancies in the proton.  In contrast, assumptions concerning
charm quark fragmentation do not strongly affect the relative rates.  In
Fig.~\ref{profus}(b) we see that the relative $q \overline q$ contribution to 
$pp$ production
is somewhat larger at $x_F \approx 0$ than in $\Sigma^- p$ production at the
same energy but this does not affect the point where $gg$ fusion ceases to
dominate $c \overline c$ production. 
The pion valence distributions are harder, allowing charm production 
at larger $x_F$ than with a baryon beam, as shown in Fig.~\ref{pifus}.  
For the pion, we use the GRV LO pion set \cite{GRVpi} with the GRV LO 94 
proton set and with the MRS D$-^\prime$ distributions we use the SMRS P2 pion
distributions \cite{SMRS}.  However, the 
valence $\overline u$ quark in the $\pi^-$ does not change the relative
importance of $q \overline q$ annihilation at 650 GeV, as seen in
Fig.~\ref{pifus}(b). 
Much lower energies are needed for the $\pi^-$ antiquark to lead to dominance
of $q \overline q$ annihilation in $c \overline c$ production.

The charm hadron distributions from PYTHIA 6.115 \cite{PYT} for the three
projectiles at 330 GeV beam energy
are also shown in Figs.~\ref{sigfus}-\ref{pifus} (c) and (d).  
The PYTHIA calculations,
based on $10^7$ events, use all default program settings along with
the GRV LO 94 parton distributions.  We note that in PYTHIA
the hyperon valence quark distributions are an average of the proton valence
distributions, $d^\Sigma_v = s^\Sigma_v = (u_v^p + d_v^p)/3$.
In (c) the $D^-$, $D_s^+$, $\Lambda_c$ and $\Xi_c^0$ $x_F$ distributions
are shown while the $D^+$, $D_s^-$, $\Sigma_c^0$ and $\Xi_c^+$ distributions
are given in (d).
The magnitude of the curves reflect the relative abundancies of charm hadrons
produced by PYTHIA.  

The Lund string fragmentation model \cite{PYT} produces charm quarks at
string endpoints.  The strings pull the charm quarks toward the
opposite endpoints, typically the beam remnants.  When the two
string endpoints are moving in the same general direction, the
charm hadron can be produced with larger longitudinal momentum
than the charm quark.  In the case where
the string invariant mass is too small for multiple particle
production, a single hadron is produced \cite{torb}, as in the $\Sigma_c^0 
(ddc)$ and $\Xi_c^0 (dsc)$ which share two valence quarks with the $\Sigma^-$.
These distributions have a minimum at $x_F \sim 0.3$ and 0.1 respectively
and a peak at $x_F \sim 0.8$, illustrating the acceleration undergone by
charm quarks by strings with small invariant mass.  The $\Lambda_c$ and
$\Xi_c^+$ are also
accelerated by string fragmentation but the effect is not as strong because 
only one valence quark is in common with the projectile.  

In contrast, with a proton beam, as shown in Fig.~\ref{profus}, 
only the $\Lambda_c$ 
shows strong forward acceleration due to the common $u$ and $d$ quarks with the
maximum in the $x_F$ distribution occurring at $x_F \approx 0.8$.  A second
peak is notable for the $\Sigma_c^0$ but the acceleration effect is weaker for 
charm-strange baryon production, presumably due to the additional mass of the
strange quark.  While meson production does not show any significant leading
behavior with a baryon projectile, the situation is reversed with the $\pi^-$
beam where $D$ and $D_s$ production is clearly forward of all charm baryon
production, produced centrally in the forward $x_F$ region.
We will make further comparisons with PYTHIA when our full model is discussed.

\section{Intrinsic Particle Production}
\paragraph{}
 
The wavefunction of a hadron in QCD can be represented as a
superposition of Fock state fluctuations, {\it e.g.}\ $\vert n_V
\rangle$, $\vert n_V g \rangle$, $\vert n_V Q \overline Q \rangle$,
\ldots components where $n_V \equiv dds$ for a $\Sigma^-$, $uud$ for a proton
and $\overline u d$ for a $\pi^-$. 
When the projectile scatters in the target, the
coherence of the Fock components is broken and the fluctuations can
hadronize either by uncorrelated fragmentation as for leading twist
production or coalescence with
spectator quarks in the wavefunction \cite{intc,BHMT}.  The
intrinsic heavy quark Fock components are generated by virtual
interactions such as $g g \rightarrow Q \overline Q$ where the
gluons couple to two or more projectile valence quarks. The
probability to produce $Q \overline Q$ fluctuations scales as
$\alpha_s^2(M_{Q \overline Q})/m_Q^2$ relative to leading-twist
production \cite{BH}. Intrinsic $Q
\overline Q$ Fock states are dominated by configurations with
equal rapidity constituents so that, unlike sea quarks generated
from a single parton, the intrinsic heavy quarks carry a large
fraction of the parent momentum \cite{intc}.
 
The frame-independent probability distribution of an $n$--particle
$c \overline c$ Fock state is
\be
\frac{dP^n_{\rm ic}}{dx_i \cdots dx_n} = N_n \alpha_s^4(M_{c \overline
c}) \frac{\delta(1-\sum_{i=1}^n x_i)}{(m_h^2 - \sum_{i=1}^n
(\widehat{m}_i^2/x_i) )^2} \, \, ,
\label{icdenom}
\ee
where $N_n$ normalizes the $|n c
\overline c \rangle$ probability, $P^n_{\rm ic}$, and $n=4$, 5 for meson and
baryon production from the $|n_V c \overline c \rangle$ 
configuration.  The delta function
conserves longitudinal momentum.  The dominant Fock configurations
are closest to the light-cone energy shell and therefore the invariant
mass, $M^2 = \sum_i \widehat{m}_i^2/ x_i$, is minimized where
$\widehat{m}_i^2 =k^2_{T,i}+m^2_i$ is the effective transverse mass
of the $i^{\rm th}$ particle and $x_i$ is the light-cone momentum
fraction.  Assuming $\langle \vec k_{T, i}^2 \rangle$ is
proportional to the square of the constituent quark mass, we choose
$\widehat{m}_q = 0.45$ GeV, $\widehat{m}_s =
0.71$ GeV, and $\widehat{m}_c = 1.8$ GeV \cite{VBH1,VBH2}.
 
The intrinsic charm production cross section for a single charm hadron from 
the $n$-particle state can be related to
$P^n_{\rm ic}$ and the inelastic $hN$ cross section by
\be
\sigma^n_{\rm ic}(hN) = P^n_{\rm ic} \sigma_{h N}^{\rm in}
\frac{\mu^2}{4 \widehat{m}_c^2} \, \, .
\label{icsign}
\ee
The factor of $\mu^2/4 \widehat{m}_c^2$ arises from the soft
interaction which breaks the coherence of the Fock state. To set the scale of
the coherence factor $\mu$ we asssume that the NA3 diffractive $J/\psi$  cross
section \cite{Badier} can be attributed to intrinsic charm.  In this experiment
the nuclear dependence of $J/\psi$ production in $\pi^- A$ interactions 
into a ``hard''
contribution with a nearly linear $A$ dependence at low $x_F$ and a
high $x_F$ ``diffractive'' contribution scaling as $A^\beta$ where $\beta =
0.77$ for pion and 0.71 for proton beams,
characteristic of soft interactions.  Then we assume that
the diffractive fraction of the production
cross section \cite{Badier} is the same for charmonium and charm
hadrons.  In Ref.\ \cite{VB}, $\mu^2 \sim 0.2$ GeV$^2$ was found, however,
calculations with more recent parton densities suggest that $\mu^2 \sim 0.1$
GeV$^2$.  We thus obtain 
$\sigma^4_{\rm ic}(\pi N) \approx 0.5$
$\mu$b and $\sigma^5_{\rm ic}(p N) \approx 0.7$ $\mu$b
at 200 GeV.  We take $P^5_{\rm ic} = 0.31$\%, as determined
from an analysis of the EMC
charm structure function data \cite{EMCic}. A recent reanalysis of
the EMC data with next-to-leading order calculations of leading twist
and intrinsic charm electroproduction is consistent with the presence of an
intrinsic charm component in the proton at large
$x_{\rm Bj}$ of $\approx 1$\%\ or less \cite{hsv}.  For simplicity,
we will always assume that the total probability for a charm quark in an
$|n_V c \overline c \rangle$ state is 0.31\% \cite{EMCic,hsv}, 
regardless of the projectile identity.

The inelastic $\Sigma^{-}p$ cross section has not been measured.  However the
total and elastic $\Lambda p$ cross sections have been parameterized for beam
momenta less than 200 GeV albeit with large statistical uncertainties.
Extrapolating these cross sections to 330 GeV, we found that $\sigma_{\Lambda
p}^{\rm in} > \sigma_{pp}^{\rm in}$ at this energy which seems unlikely.  
To be conservative, we therefore scaled
$\sigma_{pp}^{\rm in}$ to $\sigma_{\Lambda p}^{\rm in}$ at the highest measured
$\Lambda p$
energy and used the energy dependence of $\sigma_{pp}^{\rm in}$ thereafter
to set the
scale for $\Sigma^- p$ interactions at larger values of $\sqrt{s}$.

There are two ways of producing charm hadrons from intrinsic $c
\overline c$ states.  The first is by uncorrelated fragmentation, previously
discussed in Section 2.  Additionally, if the projectile has the
corresponding valence quarks, the charm quark can also hadronize
by coalescence with the valence spectators. The coalescence
mechanism thus introduces flavor correlations between the projectile
and the final-state hadrons, producing {\it e.g.}\ $\Xi_c^0$'s with a
large fraction of the $\Sigma^-$ momentum.

First we briefly discuss charm production by uncorrelated fragmentation.
If we assume that the $c$ quark fragments into a $D$ meson, the $D$
distribution is
\be
\frac{d P^{nF}_{\rm ic}}{dx_{D}} = \int dz \prod_{i=1}^n dx_i
\frac{dP^n_{\rm ic}}{dx_1 \ldots dx_n} D_{D/c}(z) \delta(x_{D} - z
x_c) \, \, ,
\label{icfrag}
\ee
These distributions
are assumed for all intrinsic charm production by uncorrelated
fragmentation with $D_{H/c}(z) = \delta(z-1)$.  
We will not use Peterson function fragmentation further in this work.

The coalescence distributions, on the other hand, are specific for
the individual charm hadrons.  It is reasonable to assume that the intrinsic
charm Fock states are fragile and can easily materialize into charm hadrons
in high-energy, low momentum transfer reactions through coalesence.   These
contributions, taken from Ref.\ \cite{intc}, do not include any binding energy
of the produced hadrons or any mass effects.
The coalescence contribution to
charm hadron production is
\be
\frac{d P^{nC}_{\rm ic}}{dx_H} = \int \prod_{i=1}^n dx_i
\frac{dP^n_{\rm ic}}{dx_1 \ldots dx_n} \delta(x_H - x_{H_1}-\cdots - 
x_{H_{n_V}}) \, \, . 
\label{iccoalD}
\ee
The coalescence function is simply a delta function combining the momentum
fractions of the quarks in the Fock state configuration that make up the
valence quarks of the final-state hadron.

We now compare and contrast $D^-$, $D^+$, $D_s^-$ and $D_s^+$ meson 
and $\Lambda_c$, 
$\Sigma_c^0$, $\Xi_c^0$ and $\Xi_c^+$ baryon production by coalescence from
$\Sigma^-$, $p$ and $\pi^-$ projectiles.  We note that not all of these hadrons
can be produced from the minimal intrinsic charm Fock state configuration,
$|n_V c \overline c \rangle$. 
However, coalescence can also occur within higher fluctuations of the
intrinsic charm Fock state.  For example, in the proton, the $D^+$ and
$\Xi_c^0$ can be produced by coalescence from $|n_V c
\overline c d \overline d \rangle$ and $|n_V c \overline c s
\overline s \rangle$ configurations.  These higher Fock state probabilities can
be obtained using earlier results \cite{VBlam,VB2}. 
In a previous study of $\psi
\psi$ production from  $|n_V c \overline c c \overline c \rangle$
states \cite{VB2} the double intrinsic charm production
probability, $P_{\rm icc}$, was determined assuming that 
all the measured $\psi \psi$ pairs
\cite{Badpi,Badp} arise from these configurations.  The resulting 
upper bound on the model,
$\sigma_{\psi \psi} = \sigma_{\rm ic}^{\psi \psi} (\pi^- N) \approx
20$ pb set by experiment \cite{Badpi}, 
requires $P_{\rm icc} \approx 4.4\%\ P_{\rm
ic}$ \cite{VB2,RV}. This value can then be used to
estimate the probability of light quark pairs in an intrinsic charm
state. We expect that the probability of additional light quark
pairs in the Fock states to be larger than $P_{\rm icc}$,
\be
P_{\rm icq} \approx \left( \frac{\widehat{m}_c}{\widehat{m}_q}
\right)^2 P_{\rm icc} \, \, ,
\label{icrat}
\ee
leading to $P_{\rm icu} = P_{\rm icd} \approx 70.4\%\ P_{\rm ic}$
and $P_{\rm ics} \approx 28.5\%\ P_{\rm ic}$.  To go to still higher
configurations, {\it e.g.}\ for $\Xi_c^+$ production from a $\pi^-$, one can
make the similar assumption that $P_{\rm icsu} = 70.4\% P_{\rm ics}$.
 
In Table~\ref{icconfig} 
we show the minimum number of partons needed in each configuration
to produce a given charm hadron.  When more than the minimal $|n_V c \overline
c \rangle$ state is necessary for coalescence to occur, the additional light
quark pairs required in the state are indicated.  While we include the eight
particle configuration necessary to produce a $\Xi_c^+$ by coalescence from
a $\pi^-$ projectile, we will confine our discussion to charm hadron
production from the minimal state and states with one additional $q \overline
q$ pair only.
\begin{table}
\begin{center}
\begin{tabular}{|c|c|c|c|} \hline
Particle & $\Sigma^- (dds)$ & $p(uud)$ & $\pi^-(\overline u d)$ \\ \hline
$D^- (d \overline c)$ & 5 & 5 & 4 \\ \hline
$D^+ (\overline d c)$  & $7(d \overline d)$ & $7(d \overline d)$ & 
$6(d \overline d)$ \\ \hline
$\Lambda_c (udc)$ &  $7(u \overline u)$ & 5 & $6(u \overline u)$ \\ \hline
$\Sigma_c^0(ddc)$ & 5 & $7(d \overline d)$ &  $6(d \overline d)$ \\ \hline
$D_s^- (s \overline c)$ & 5 & $7(s \overline s)$ & $6(s \overline s)$ \\ \hline
$D_s^+ (\overline s c)$ & $7(s \overline s)$ & $7(s \overline s)$ & 
$6(s \overline s)$ \\ \hline 
$\Xi_c^0(dsc)$ & 5 &  $7(s \overline s)$ &  $6(s \overline s)$ \\ \hline
$\Xi_c^+(usc)$ & $7(u \overline u)$ & $7(s \overline s)$ & $8(s \overline s u
\overline u)$ \\ \hline
\end{tabular}
\end{center}
\caption[]{The lowest number of partons needed in an intrinsic charm Fock state
configuration for the charm particle to be produced by coalescence.  Note that
4 and 5 correspond to the minimal $|n_V c \overline c \rangle$ configuration
while the higher states refer to $|n_V c \overline c d \overline d \rangle$
{\it etc.}}
\label{icconfig}
\end{table}

The total intrinsic charm contribution to charm hadron production is a
combination of uncorrelated fragmentation and coalescence.  In previous works
\cite{VB,VBlam,VBH1,VBH2} only production by uncorrelated fragmentation
from the minimal $|n_V c \overline c \rangle$ states and coalescence from the
minimum Fock state configuration was considered.  This was 
because a significant leading effect is present only in the minimal
configuration, {\it i.e.} there is no difference between $D^+$ and $D^-$ mesons
produced from $|n_V c \overline c d \overline d \rangle$ states.  Also, as more
partons are included in the Fock state, the coalescence
distributions soften and approach the fragmentation distributions,
eventually producing charm hadrons with less momentum
than uncorrelated fragmentation from the minimal $c \overline c$
state if a sufficient number of $q \overline q$ pairs are included.
There is then no longer any advantage to introducing more light
quark pairs into the configuration---the relative probability will
decrease while the potential gain in momentum is not significant.
However, if some fraction of the final-state hadrons are assumed to be produced
from higher Fock configurations, then all possible final-states from those
configurations should also be included.  Therefore in this paper, we consider
production by fragmentation and coalescence from the minimal state and the next
higher states with $u \overline u$,  $d \overline d$ and $s \overline s$ pairs.

\begin{figure}[htb]
\setlength{\epsfxsize=0.95\textwidth}
\setlength{\epsfysize=0.5\textheight}
\centerline{\epsffile{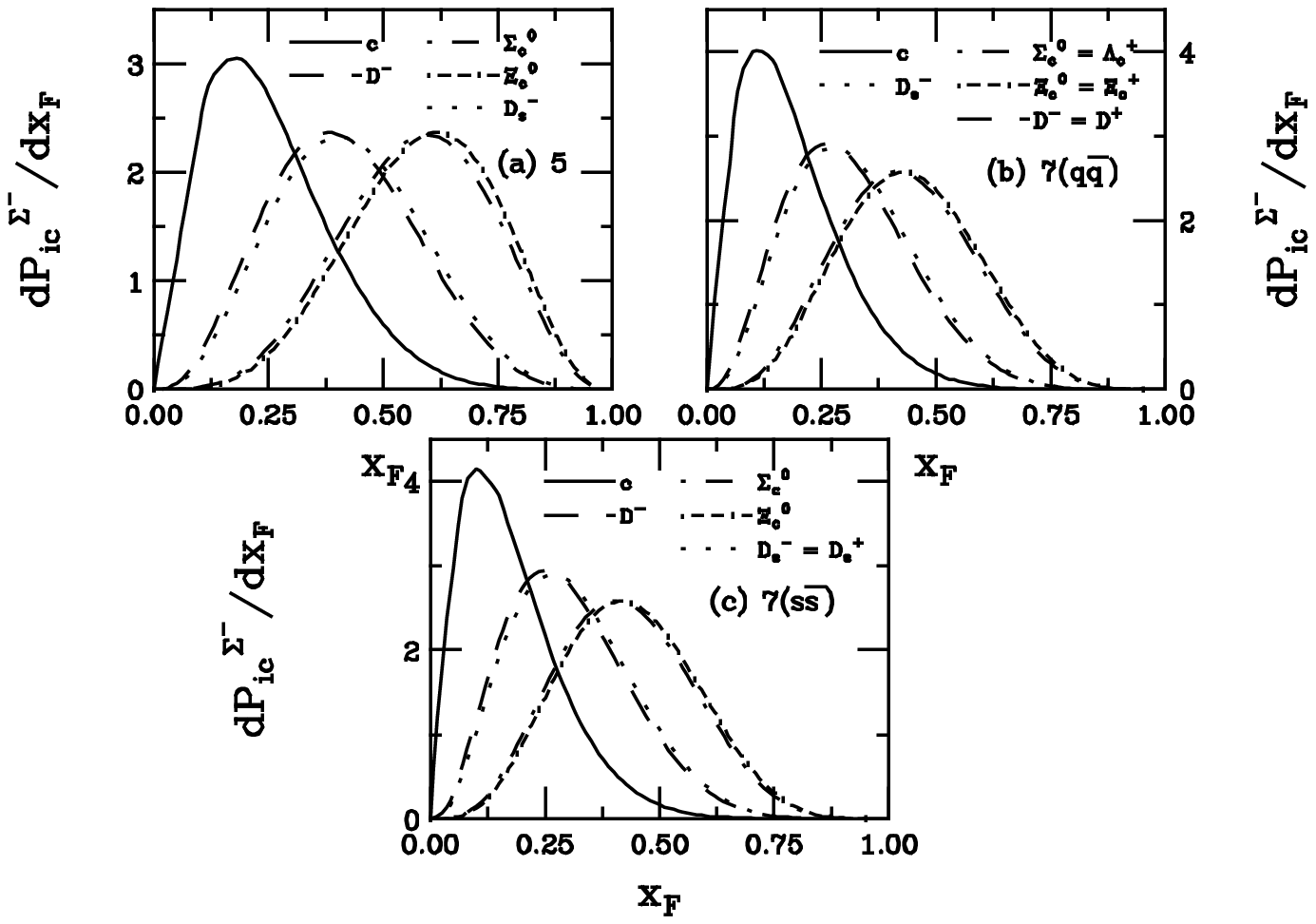}}
\caption[]{ Charm hadron production in the intrinsic charm model with a
$\Sigma^-$ projectile.  The probability distributions, $(1/P^n_{\rm 
ic})(dP^n_{\rm ic}/dx_H)$, for uncorrelated fragmentation and coalescence are
given for the minimal 5-particle Fock state (a) and for the 7-particle Fock 
states with light quarks $q=u$, $d$ (b) and with strange quarks (c).  The solid
curve in each case is the charm quark distribution which also serves as the
hadron distribution for independent fragmentation.  The other curves are
the probability distributions for hadron production by coalescence, including:
$D^-$ (dashed), $\Sigma_c^0$ (dot-dashed), $\Xi_c^0$ (dot-dash-dashed)
and $D_s^-$ (dotted).  If the shape of the probability distribution is the
same for any two hadrons (such as the $\Sigma_c^0$ and the $\Lambda_c^+$ in 
(b)) in a configuration, it is indicated.}
\label{sigic}
\end{figure}

\begin{figure}[htb]
\setlength{\epsfxsize=0.95\textwidth}
\setlength{\epsfysize=0.5\textheight}
\centerline{\epsffile{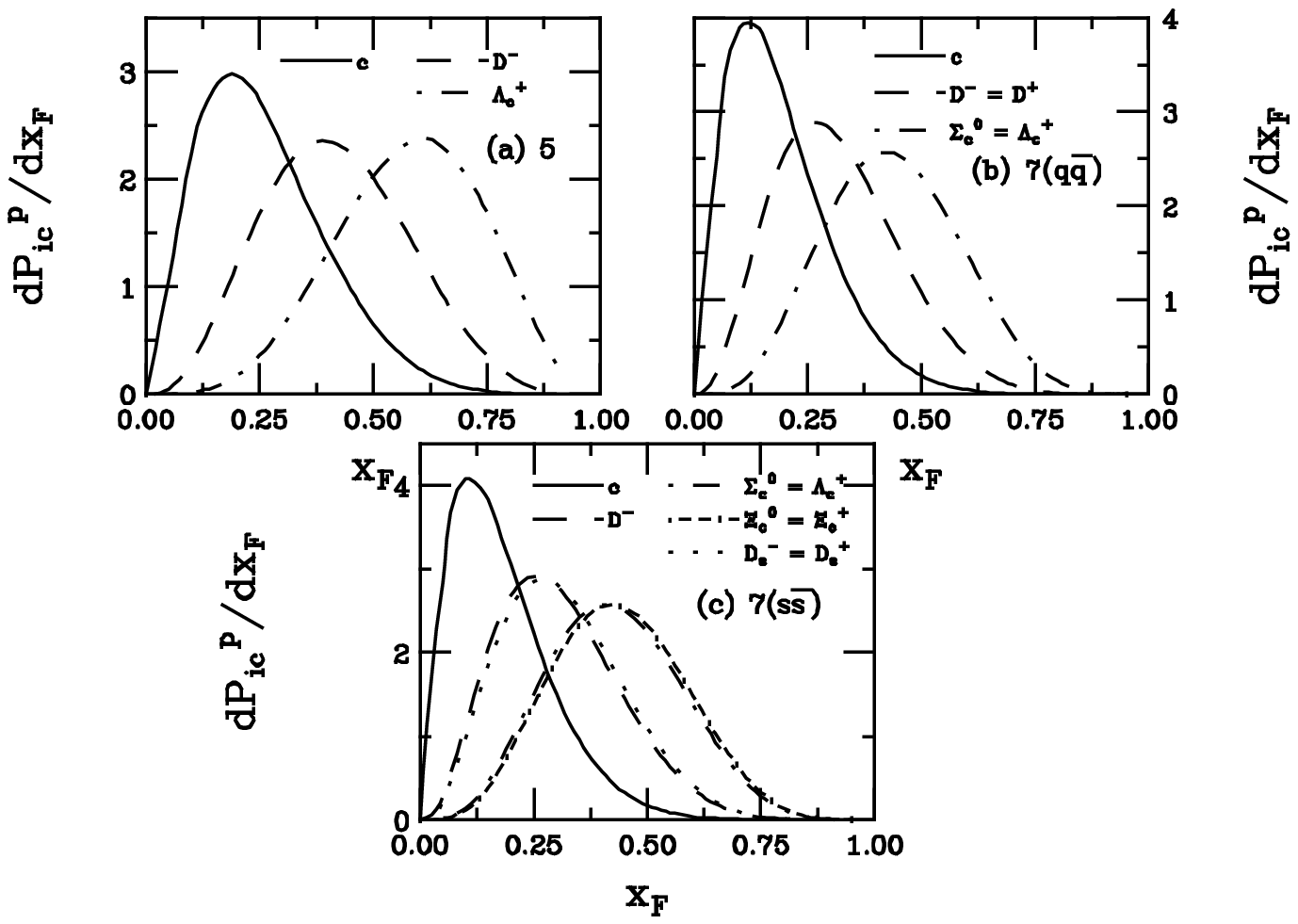}}
\caption[]{ Charm hadron production in the intrinsic charm model with a
proton projectile.  The probability distributions, $(1/P^n_{\rm 
ic})(dP^n_{\rm ic}/dx_H)$, for uncorrelated fragmentation and coalescence are
given for the minimal 5-particle Fock state (a) and for the 7-particle Fock 
states with light quarks $q=u$, $d$ (b) and with strange quarks (c).  The solid
curve in each case is the charm quark distribution which also serves as the
hadron distribution for independent fragmentation.  The other curves are
the probability distributions for hadron production by coalescence, including:
$D^-$ (dashed), $\Lambda_c^+$ (dot-dashed), $\Xi_c^0$ (dot-dash-dashed)
and $D_s^-$ (dotted).  If the shape of the probability distribution is the
same for any two hadrons (such as the $\Sigma_c^0$ and the $\Lambda_c^+$ in 
(b)) in a configuration, it is indicated.}
\label{proic}
\end{figure}

\begin{figure}[htb]
\setlength{\epsfxsize=0.95\textwidth}
\setlength{\epsfysize=0.5\textheight}
\centerline{\epsffile{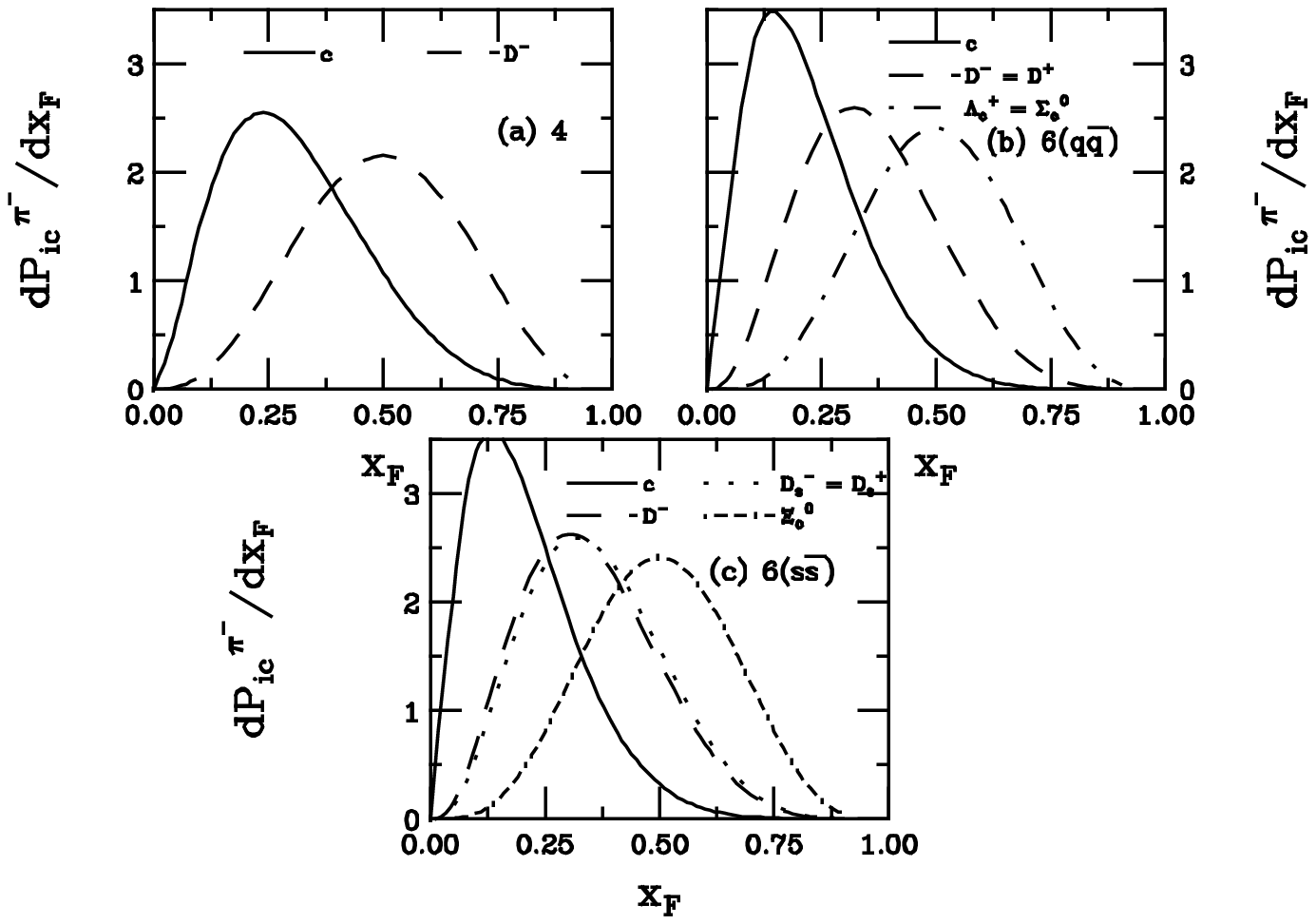}}
\caption[]{ Charm hadron production in the intrinsic charm model with a
$\pi^-$ projectile.  The probability distributions, $(1/P^n_{\rm 
ic})(dP^n_{\rm ic}/dx_H)$, for uncorrelated fragmentation and coalescence are
given for the minimal 5-particle Fock state (a) and for the 7-particle Fock 
states with light quarks $q=u$, $d$ (b) and with strange quarks (c).  The solid
curve in each case is the charm quark distribution which also serves as the
hadron distribution for independent fragmentation.  The other curves are
the probability distributions for hadron production by coalescence, including:
$D^-$ (dashed), $\Lambda_c^+$ (dot-dashed), $\Xi_c^0$ (dot-dash-dashed)
and $D_s^-$ (dotted).  If the shape of the probability distribution is the
same for any two hadrons (such as the $\Sigma_c^0$ and the $\Lambda_c^+$ in
(b)) in a configuration, it is indicated.}
\label{piic}
\end{figure}

The probability distributions, $(1/P^n_{\rm ic})(dP^n_{\rm ic}/dx_H)$, 
are given in Figs.~\ref{sigic}-\ref{piic} 
for $\Sigma^-$, $p$ and $\pi^-$ projectiles respectively.  It
is clear from Fig.~\ref{sigic} 
that the $\Sigma^-$ projectile allows the greatest
coalescence production of charm hadrons from the minimal Fock configuration,
Fig.~\ref{sigic}(a).  
The charm baryons are quite fast, taking more than 50\% of the
projectile momentum.  The difference between charm and charm-strange hadron
production is very small due to the strange and light quark mass difference.
Because the strange quark is more massive, it carries a somewhat larger
fraction of the $\Sigma^-$ momentum than the light quarks, resulting in a
slightly larger average momentum for the $\Xi_c^0$ and the $D_s^-$ relative
to the $\Sigma_c^0$ and $D^-$, on the order of 3--4\% 
as can be seen in Table~\ref{icavexf}.  
The $c$ quark distribution itself, leading to
uncorrelated fragmentation, carries $\approx 25$\% of the projectile momentum
in the minimal state.  This is reduced by $\approx 35$\% in the seven-particle
Fock configurations.  In this model, the $c$ and $\overline c$ probability
distributions are identical.
We note that these higher configurations can produce, for example,
$\Sigma_c^0$ and $\Lambda_c$ baryons from $|n_V c \overline c d
\overline d \rangle$ and $|n_V c \overline c u \overline u \rangle$ states
respectively with the same probability distribution, shown in
Fig.~\ref{sigic}(b), but 
not necessarily with the same relative probability, as we will show shortly.
Introducing an $s \overline s$ pair to the 7-particle configuration reduces
the average momentum of the final state hadron by $\approx 2$\% over the
average in the 7-particle configurations with lighter $q \overline q$ pairs.
In addition to the reduction of the average momentum of the $c$ quark in the
higher configurations, the final-state charm hadron momentum from this
configuration is reduced as well, suggesting that no more significant
contribution to the overall momentum of the final hadron will be obtained
by including yet higher Fock configurations.

While fewer charm hadrons can be directly produced from the minimal
configuration of a proton projectile, as evident from Fig.~\ref{proic}, 
their average 
momentum is somewhat higher than the $\Sigma^-$ due to the abscence of the
strange valence quark.  However, this only affects the final-state average
momentum by 1-2\%.  Final-state charm hadrons from a pion projectile, shown in
Fig.~\ref{piic} have, on
average, 20\% more momentum than from a baryon projectile because the total
velocity is shared between fewer initial partons.  Note also that mesons from
a four-particle Fock configuration and baryons from a six-particle Fock state
each receive half of the projectile momentum.

\begin{table}
\begin{center}
\begin{tabular}{|c|c|c|c|c|} \hline
State & Particle & $\Sigma^- (n_V=dds)$ & $p(n_V=uud)$ & 
 $\pi^-(n_V=\overline u d)$ \\ \hline
$|n_V c \overline c \rangle$ & $c$ & 0.251 & 0.256 & 0.308 \\ \hline
 '' & $D^- (d \overline c)$ & 0.41 & 0.419 & 0.499 \\ \hline
 '' & $\Lambda_c (udc)$ & - & 0.58 & - \\ \hline
 '' & $\Sigma_c^0(ddc)$ & 0.573 & - & -  \\ \hline
 '' & $D_s^- (s \overline c)$ & 0.427 & - & - \\ \hline
 '' & $\Xi_c^0(dsc)$ & 0.59 & - & - \\ \hline
$|n_V c \overline c q \overline q \rangle$ & $c$ & 0.185 & 0.188 & 0.219 
\\ \hline
 '' & $D^- (d \overline c) = D^+ (\overline d c)$ & 0.31 & 0.314 & 0.359 
\\ \hline
 '' & $\Lambda_c (udc)$ & 0.433 & 0.438 & 0.5 \\ \hline
 '' & $\Sigma_c^0(ddc)$ & 0.433 & 0.438 & 0.5  \\ \hline
 '' & $D_s^- (s \overline c)$ & 0.32 & - & - \\ \hline
 '' & $\Xi_c^0(dsc) = \Xi_c^+(usc)$ & 0.444 & - & - \\ \hline
$|n_V c \overline c s \overline s \rangle$ & $c$ & 0.179 & 0.181 & 0.211 
\\ \hline
 '' & $D^- (d \overline c)$ & 0.302 & 0.306 & 0.349 \\ \hline
 '' & $\Lambda_c (udc)$ & - & 0.429 & - \\ \hline
 '' & $\Sigma_c^0(ddc)$ & 0.424 & - & -  \\ \hline
 '' & $D_s^- (s \overline c) = D_s^+ (\overline s c)$ & 0.312 & 0.316 & 0.361 
\\ \hline
 '' & $\Xi_c^0(dsc)$ & 0.434 & 0.439 & 0.5 \\ \hline
 '' & $\Xi_c^+(usc)$ & 0.434 & - & - \\ \hline
\end{tabular}
\end{center}
\caption[]{The average value of $x_F$ for charm particles produced by 
coalescence from $\Sigma^-$, $p$ and $\pi^-$ projectiles in $|n_V c \overline c
\rangle$, $|n_V c \overline c q \overline q \rangle$ and $|n_V c \overline c s
\overline s \rangle$ states.  In this case, $q \overline q = u \overline u$, $d
\overline d$.}
\label{icavexf}
\end{table}

\section{Model Predictions}
 
We now turn to specific predictions of our model. 
We begin with the $x_F$ distribution of the final-state charm hadrons.
The $x_F$ distribution for final-state hadron $H$ is the sum
of the leading-twist fusion and intrinsic charm components,
\be
\frac{d\sigma^H_{hN}}{dx_F} = \frac{d\sigma^H_{\rm lt}}{dx_F} +
\frac{d\sigma^H_{\rm ic}}{dx_F} \, \, ,
\label{tcmodel}
\ee
where $d\sigma^H_{\rm ic}/dx_F$ is related to $dP^H/dx_F$ by
\be
\frac{d\sigma^H_{\rm ic}}{dx_F} = \sigma_{h N}^{\rm in}
\frac{\mu^2}{4 \widehat{m}_c^2} \frac{dP_H}{dx_F} \, \, .
\label{iccross}
\ee
The probability distribution is the sum of all contributions from the $|n_V c
\overline c \rangle$ and the $|n_V c
\overline c q \overline q \rangle$ configurations with $q = u$, $d$, and $s$
and includes uncorrelated fragmentation and coalescence when appropriate, as
described below.  We use the
same fragmentation function, either the delta or Peterson function, to
calculate uncorrelated fragmentation in 
both leading twist fusion and intrinsic charm.  In this section, we use only
the delta function.

Since experimental information on the relative rates of charm hadron production
is incomplete, we assume that all the lowest lying charm hadrons produced
by uncorrelated fragmentation have equal probability in both leading-twist
fusion and intrinsic charm.  There are 10 charm hadrons---and 
the same number of
anticharm hadrons---if excited charm hadrons such as $D^*$ and
$\Lambda_c^+(2593)$ are excluded.  Therefore the
probability distribution for uncorrelated fragmentation into each of these
hadrons is 10\% of the total probability.  As can be seen in Table 1, only a
fraction of the possible final-state hadrons can be produced by coalescence.
We use a simple counting scheme to arrive at the coalescence probability
which enhances the production of leading charm at large $x_F$.
We note that the combined probability of fragmentation and coalescence of all
charm hadrons cannot exceed the total production probability 
of the Fock state configuration.  Thus when a particular final-state hadron can
be produced both by uncorrelated fragmentation and coalescence, we multiply
the sum of the fragmentation and coalescence probabilities by 0.5 to keep the
total probability fixed.

As a concrete example of how the total probability distributions of charm
hadron production from the intrinsic charm model is calculated, 
we will describe $D^+$ and $D^-$ production from the $\Sigma^-$ beam
in our model in detail.  The full complement of equations
for all the final-state charm hadrons from $\Sigma^-$, $p$ and $\pi^-$
projectiles considered in this work can be found in the appendix.  In the
$|ddsc \overline c \rangle$ configuration, there are four
final-state hadrons with a valence $c$ quark (2$\Xi_c^0$, $\Sigma_c^0$ and
$J/\psi$) and also four final-state hadrons with a valence $\overline c$ quark
(2$D^-$, $D_s^-$ and $J/\psi$).  Note that the $J/\psi$ has been 
counted in each
category.  The $D^-$ is then produced by coalescence with 50\% of the total
coalescence probability for hadrons with a valence $\overline c$ as well as by
uncorrelated fragmentation of the $\overline c$ while the
$D^+$ is only produced by uncorrelated fragmentation from this state.  The 
probability distributions from this minimal Fock configuration are then
\be \frac{dP_{D^-}^5}{dx_F} & = & \frac{1}{2} \left( \frac{1}{10} \frac{dP_{\rm
ic}^{5F}}{dx_F} + \frac{1}{2} \frac{dP_{\rm ic}^{5C}}{dx_F} \right) \\
 \frac{dP_{D^+}^5}{dx_F} & = & \frac{1}{10} \frac{dP_{\rm ic}^{5F}}{dx_F}
\label{dpdmmin}
\ee
where $F$ refers to uncorrelated fragmentation and $C$ to coalescence into the
specific final-state with the associated probability distribution, shown in
Fig.~\ref{sigic}(a).  
The $|ddsc \overline c q \overline q \rangle$ configurations where
$q \overline q = u \overline u$, $d \overline d$ and $s \overline s$ allow
coalescence production of eight final-state hadrons with a valence $c$ and five
final-state hadrons with a valence $\overline c$ in each case.  
We will discuss $D^+$ and
$D^-$ from each of these configurations in turn.  When $q \overline q = u
\overline u$, the possible hadrons produced by coalescence are: 2$\Xi_c^+$,
$\Xi_c^0$, 2$\Lambda_c^+$, $\Sigma_c^0$, $D^0$ and $J/\psi$ with a valence $c$
and 2$D^-$, $D_s^-$, $\overline D^0$ and $J/\psi$ with a valence $\overline c$.
A final-state $D^+$ can be produced by coalescence from the $d \overline d$
configuration in one of the eight possible
final-state hadrons with a valence $c$ quark (3$\Sigma_c^0$, 3$\Xi_c^0$, $D^+$
and $J/\psi$) while the $D^-$ is produced by coalescence in three out of five
combinations (3$D^-$, $D_s^-$ and $J/\psi$) with a valence $\overline c$.  
The $s \overline s$ configuration yields
no $D^+$ by coalescence---4$\Xi_c^0$, $\Sigma_c^0$, $\Omega_c^0(ssc)$, 
$D_s^+$ and 
$J/\psi$ are allowed---while 2$D^-$ are allowed out of 5 possible hadrons 
with valence $\overline c$
quarks---2$D^-$, 2$D_s^-$ and $J/\psi$.  Finally, the total intrinsic
charm probability distribution for these mesons is:
\be \frac{dP_{D^-}}{dx_F} & = & \frac{1}{2} \left( \frac{1}{10} \frac{dP_{\rm
ic}^{5F}}{dx_F} + \frac{1}{2} \frac{dP_{\rm ic}^{5C}}{dx_F} \right) + 
 \frac{1}{2} \left( \frac{1}{10} \frac{dP_{\rm
icu}^{7F}}{dx_F} + \frac{2}{5} \frac{dP_{\rm icu}^{7C}}{dx_F} \right) \nonumber
\label{dmfull} \\ &   & + \, \frac{1}{2} \left( \frac{1}{10} \frac{dP_{\rm
icd}^{7F}}{dx_F} + \frac{3}{5} \frac{dP_{\rm icd}^{7C}}{dx_F} \right) +
 \frac{1}{2} \left( \frac{1}{10} \frac{dP_{\rm
ics}^{7F}}{dx_F} + \frac{2}{5} \frac{dP_{\rm ics}^{7C}}{dx_F} \right) \\
 \frac{dP_{D^+}}{dx_F} & = & \frac{1}{10} \frac{dP_{\rm ic}^{5F}}{dx_F} +
 \frac{1}{10} \frac{dP_{\rm icu}^{7F}}{dx_F} + \frac{1}{2} \left( \frac{1}{10}
\frac{dP_{\rm icd}^{7F}}{dx_F} + \frac{1}{8} \frac{dP_{\rm icd}^{7C}}{dx_F} 
\right) + \frac{1}{10} \frac{dP_{\rm ics}^{7F}}{dx_F} \, \, . 
\label{dpfull}
\ee
Lastly we note that only fragmentation from the minimal Fock 
state was included along with coalescence from the lowest possible
configuration in earlier work \cite{VB,VBlam,VBH2}.  This corresponds to the
first term of the $D^-$ probability distribution while the $D^+$ distribution
would be proportional to $0.5 ((1/10) dP_{\rm ic}^{5F}/dx_F + (1/8) dP_{\rm
icd}^{7C}/dx_F)$. 
 
We must also account for the fact that most of the data are taken on
nuclear targets.  In this case, the model assumes a linear $A$
dependence for leading-twist fusion and an $A^\alpha$ dependence for
the intrinsic charm component \cite{Badier} where $\alpha = 0.77$ for pions
and 0.71 for protons (and $\Sigma^-$)
\be
\frac{d\sigma^H_{hA}}{dx_F} = A \frac{d\sigma^H_{\rm lt}}{dx_F} +
A^\alpha \frac{d\sigma^H_{\rm ic}}{dx_F} \, \, ,
\label{tcadep}
\ee
This $A$ dependence is
included in the calculations. The
intrinsic charm contribution to the longitudinal momentum
distributions per nucleon is thus reduced for nuclear targets.

\begin{figure}[htb]
\setlength{\epsfxsize=0.95\textwidth}
\setlength{\epsfysize=0.5\textheight}
\centerline{\epsffile{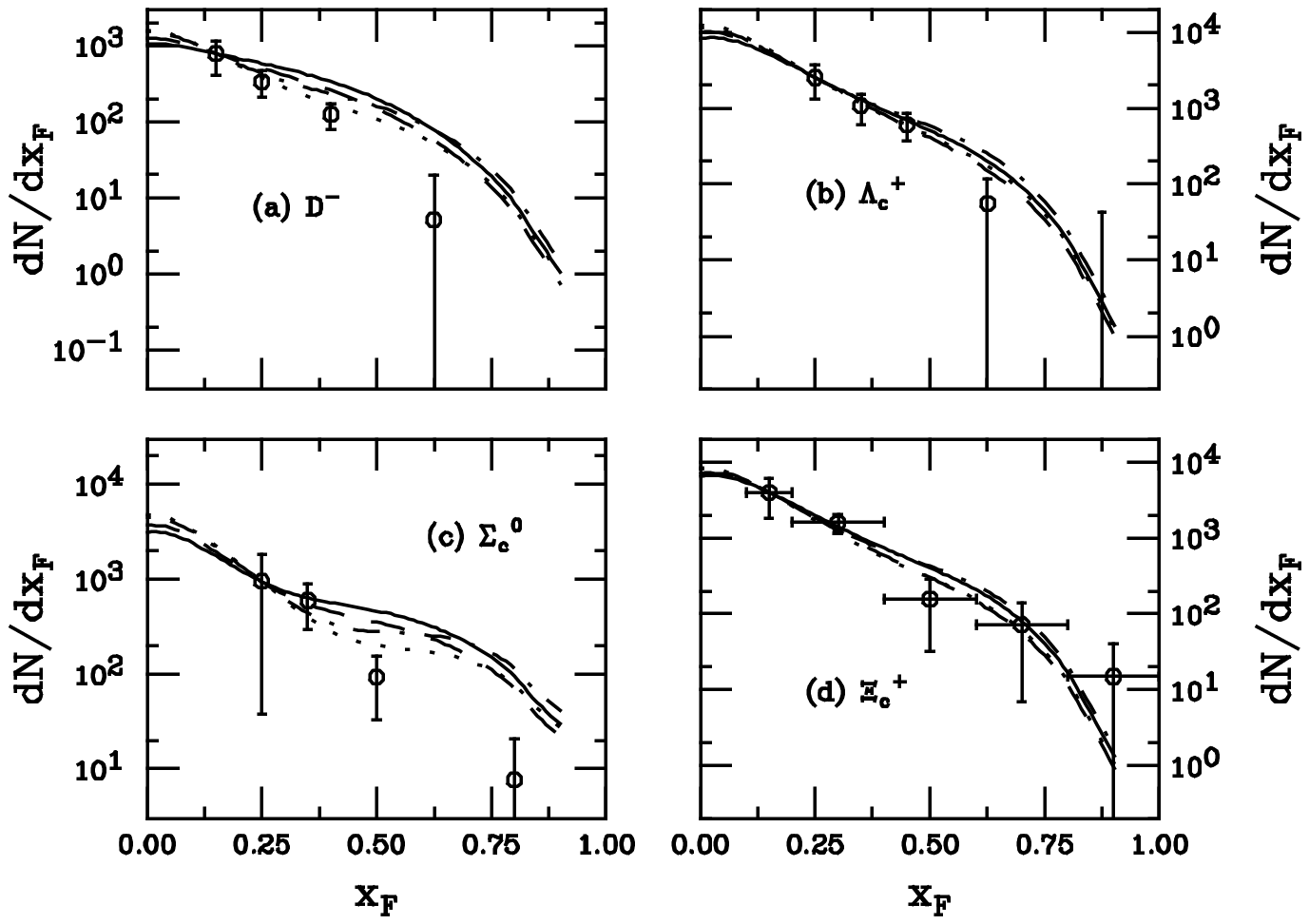}}
\caption[]{ Model predictions are compared to the $\Sigma^- A$
data of Ref.~\cite{wa89}
for (a) $D^-$, (b) $\Lambda_c^+$, (c) $\Sigma_c^0$ and (d) $\Xi_c^+$.
The solid and dashed curves represent our full model, with the intrinsic charm
probability distributions given in eqs.~(\ref{sigprobdm})-(\ref{sigprobxp}) 
for carbon and copper targets
respectively.  The dot-dashed and dotted curves contrast the results 
for carbon and copper targets respectively with the
simplified model which considers only fragmentation from the minimal Fock state
and coalescence only from the state with the minimum number of partons
necessary to produce it.}
\label{dat330}
\end{figure}

We now compare the model calculations, both the full model of
eqs.~(\ref{dmfull}) and (\ref{dpfull}) and the simpler model used previously, 
to the WA89 data
\cite{wa89} on carbon and copper targets in Fig.~\ref{dat330}.  Since the data
are unnormalized, we have normalized all curves to the first data point.
The dot-dashed and dotted curves are results with the previous simplified
model \cite{VB,VBlam,VBH2} on carbon
and copper targets respectively.  The full model is illustrated in the solid
and dashed curves for the same targets.  The agreement with the data is quite
reasonable given both the low statistics of the data and our normalization to
the first data point rather than fitting the normalization to the data.
The differences in the model distributions are most obvious for the
$\Sigma_c^0$, shown in Fig.~\ref{dat330}(c).  The simpler model emphasizes the
coalescence production from the $|n_V c \overline c \rangle$ state only.  As
can be seen from Fig.~\ref{sigic} and Table~\ref{icavexf}, the average $x_F$
of the coalescence distribution is more than a factor of two larger than that
of a $\Sigma_c^0$ production by independent fragmentation of a $c$ quark,
producing a shoulder in the $x_F$ distribution, particularly for the carbon
target (dot-dashed curve).  Because the $\Sigma_c^0$ is produced by coalescence
with $\approx 30$\% less average momentum from the 7-particle Fock states, the
intermediate $x_F$ region is partially filled in, resulting in a smoother $x_F$
distribution even though the probability is reduced for the higher Fock states.
Similar results can be seen for the other charm hadrons in Fig.~\ref{dat330}.

Note that the model results are in much better agreement with the data than the
PYTHIA simulations at the same energy with the default settings, shown in
Fig.~\ref{sigfus}(c) and (d).  If the PYTHIA predictions are superimposed on
Fig.~\ref{dat330} with the same normalization as our model, the PYTHIA results
would considerably exceed the data at large $x_F$ for the charm baryons.  In
particular, since the $c$ quark is pulled forward by a valence $dd$ diquark,
the $\Sigma_c^0$ rate from PYTHIA at $x_F \approx 0.8$ would exceed the data
by nearly four orders of magnitude.  The differences in the results are also
obvious in Table~\ref{avexffull} where the average $x_F$ of all the model
distributions on a copper target are compared to PYTHIA calculations with a
proton target at the same energy.

\begin{table}
\begin{center}
\begin{tabular}{|c|c|c|c|c|} \hline
\multicolumn{5}{|c|}{This Model} \\ \hline 
Particle & $\Sigma^-$Cu (330) & $\Sigma^-$Cu (650) & $\pi^-$Cu (650) & $p$Cu
(800) \\ \hline 
$D^- (d \overline c)$ & 0.192 & 0.147 & 0.169 & 0.120 \\ \hline
$D^+ (\overline d c)$ & 0.133 & 0.112 & 0.152 & 0.107 \\ \hline
$\Lambda_c (udc)$ & 0.145 & 0.118 & 0.154 & 0.146 \\ \hline
$\Sigma_c^0(ddc)$ & 0.187 & 0.140 & 0.154 & 0.107 \\ \hline
$D_s^- (s \overline c)$ & 0.165 & 0.129 & 0.151 & 0.106 \\ \hline
$D_s^+ (\overline s c)$ & 0.132 & 0.111 & 0.151 & 0.105 \\ \hline
$\Xi_c^0(dsc)$ & 0.221 & 0.160 & 0.151 & 0.106  \\ \hline
$\Xi_c^+(usc)$ & 0.160 & 0.126 & 0.150 & 0.106  \\ \hline \hline
\multicolumn{5}{|c|}{PYTHIA} \\ \hline
Particle & $\Sigma^- p$ (330) & $\Sigma^- p$ (650) & $\pi^- p$
(650) & $pp$ (800) \\ \hline
$D^- (d \overline c)$ & 0.14 & 0.126 & 0.254 & 0.113 \\ \hline
$D^+ (\overline d c)$ & 0.18 & 0.159 & 0.173 & 0.160 \\ \hline
$\Lambda_c (udc)$ & 0.54 & 0.468 & 0.153 & 0.604 \\ \hline
$\Sigma_c^0(ddc)$ & 0.72 & 0.707 & 0.146 & 0.35 \\ \hline
$D_s^- (s \overline c)$ & 0.155 & 0.139 & 0.172 & 0.097 \\ \hline
$D_s^+ (\overline s c)$ & 0.171 & 0.154 & 0.16 & 0.153 \\ \hline
$\Xi_c^0(dsc)$ & 0.76 & 0.767 & 0.157 & 0.123 \\ \hline
$\Xi_c^+(usc)$ & 0.55 & 0.477 & 0.156 & 0.155 \\ \hline
\end{tabular}
\end{center}
\caption[]{The average value of $x_F$ for charm particles produced in the full
model for $\Sigma^-$, $p$ and $\pi^-$ projectiles 
on a copper target.  The model
results are given at 330 GeV and 650 GeV for $\Sigma^-$Cu interactions,
650 GeV for $\pi^-$Cu interactions and 800 GeV for $p$Cu interactions. For
comparison, the average from PYTHIA at $x_F>0$ in each case is also shown.}
\label{avexffull}
\end{table}

Another way to quantify leading charm production is through the asymmetry
between leading and nonleading charm.  The asymmetry is defined as
\be
A(x_F) = \frac{d\sigma_L/dx_F - d\sigma_{NL}/dx_F}{d\sigma_L/dx_F +
d\sigma_{NL}/dx_F} \, \, 
\label{asymdef}
\ee
where $L$ represents the leading and $NL$ the nonleading charm hadron.
High statistics data has previously been available only from $\pi^-$ beams
where a significant enhancement of $D^-$ over $D^+$ production was seen at $x_F
> 0.3$ \cite{E791,WA82,E7692}, in qualitative agreement with the intrinsic
charm calculation of Ref.~\cite{VB}.  The model \cite{VBlam}
also correctly predicted the
symmetric production of $D_s^-$ and $D_s^+$ mesons and $\Lambda_c^+$ and 
$\overline \Lambda_c^+$ baryons by $\pi^-$ beams \cite{E791lam,Bar2,Adam2}.

\begin{figure}[htb]
\setlength{\epsfxsize=0.95\textwidth}
\setlength{\epsfysize=0.5\textheight}
\centerline{\epsffile{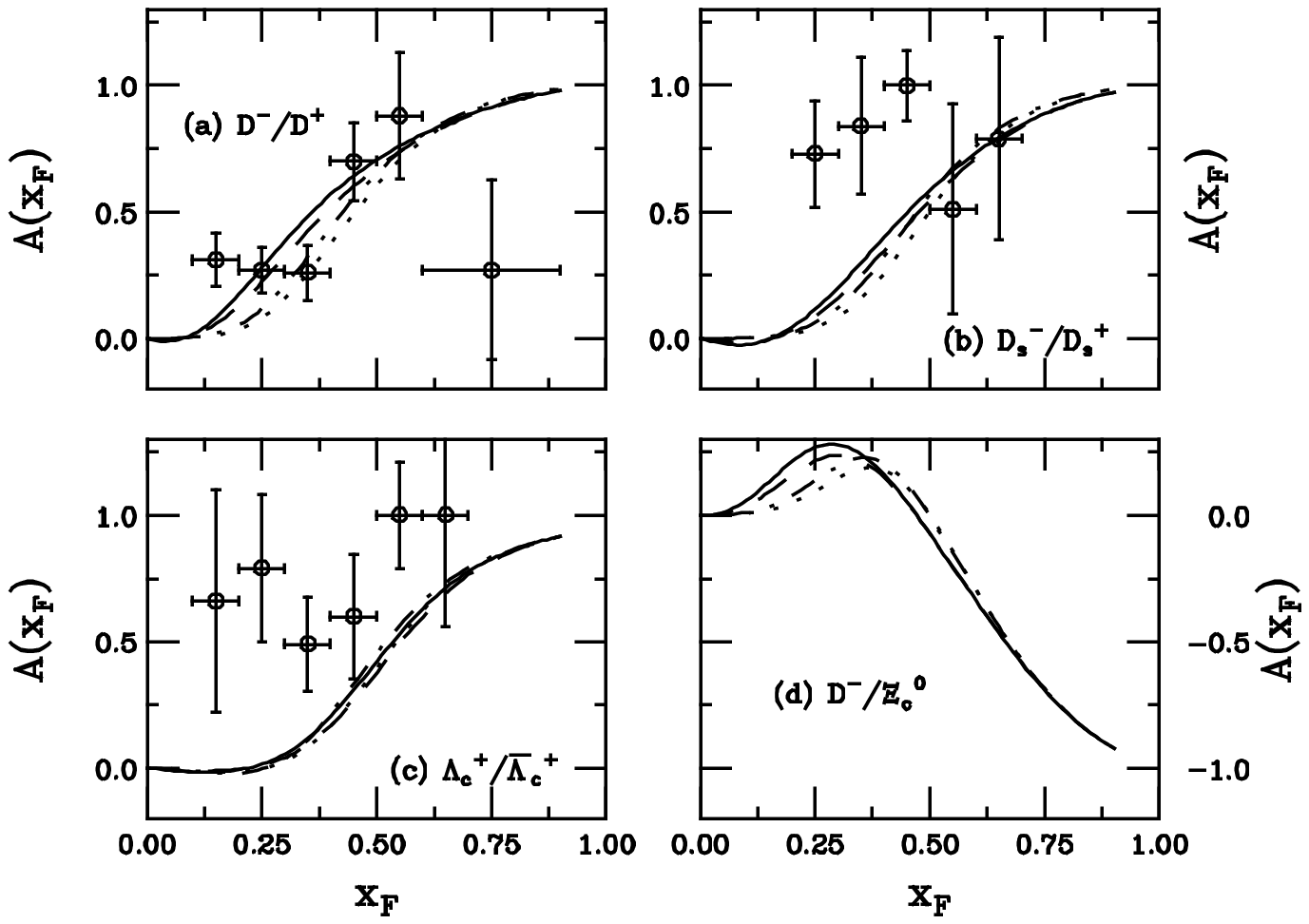}}
\caption[]{ Model predictions are compared to the $\Sigma^- A$
data of Ref.~\cite{wa89asym}
for the following asymmetries: (a) $D^-/D^+$, (b) $D_s^-/D_s^+$ and (c)
$\Lambda_c^+/\overline \Lambda_c^+$, as well as 
our prediction for the (d) $D^-/\Xi_c^0$ asymmetry.
The solid and dashed curves represent our full model, with the intrinsic charm
probability distributions given in eqs.~(\ref{sigprobdm})-(\ref{sigprobxp}) 
for carbon and copper targets
respectively.  The dot-dashed and dotted curves for carbon and copper targets
respectively contrast the results with the
simplified model which considers only fragmentation from the minimal Fock state
and coalescence only from the state with the minimum number of partons
necessary to produce it.}
\label{asym330}
\end{figure}

Statistics are unfortunately limited on charm production by baryon beams.
Recently the WA89 collaboration has presented the $D^-/D^+$, $D_s^-/D_s^+$
and $\Lambda_c^+/\overline \Lambda_c^+$ asymmetries from their $\Sigma^-$ data
\cite{wa89asym}.  In Fig.~\ref{asym330} we compare our calculations with both
models to this data as well as show a prediction for the asymmetry between the
$D^-$ and $\Xi_c^0$, both of which are produced from the partons of the minimal
Fock configuration.  The full model gives a larger asymmetry betwen $D^-$ and
$D^+$ at low $x_F$ than the simpler assumptions of previous work
\cite{VB,VBlam} because $D^-$ production at intermediate $x_F$ is enhanced by
coalescence production from the 7-particle configurations, see also
Fig.~\ref{dat330}(a).  Our results with the full model are in qualitative
agreement with the data, shown in Fig.~\ref{asym330}(a).  The measured
$D_s^-/D_s^+$ and $\Lambda_c^+/\overline \Lambda_c^+$ asymmetries are larger
than our predictions at intermediate $x_F$.  The probability distribution for
$\overline \Lambda_c^+$ in our model is
\be
 \frac{dP_{\overline \Lambda_c^+}}{dx_F} = \frac{1}{10} 
\frac{dP_{\rm ic}^{4F}}{dx_F} +
 \frac{1}{10} \frac{dP_{\rm icu}^{6F}}{dx_F} + \frac{1}{10} \frac{dP_{\rm 
icd}^{6F}}{dx_F} + \frac{1}{10} \frac{dP_{\rm ics}^{6F}}{dx_F} \, \, . 
\label{lcbdist}
\ee
Some of the discrepancies between the model and the data may arise from 
the relatively low statistics of the
$D_s$ and $\Lambda_c$ measurements.  Our model is also quite crude in overall
normalization for the different final states since we assume that all
final-state hadrons are produced by independent fragmentation with the same
probability.  Not enough high statistics data exist yet for us to use 
experimental absolute
production rates as a guide.  The asymmetry between $D^-$ and $\Xi_c^0$ is
interesting because the $|dds c \overline c \rangle$ state of the $\Sigma^-$
can be thought of as a virtual $D^- \Xi_c^0$ fluctuation, as has been suggested
for proton fluctuations into $K^+ \Lambda$ \cite{MB,SPetal} and $D^-
\Lambda_c^+$ \cite{SPetal}.  The $D^-/\Xi_c^0$ is positive at first since the
$D^-$ $x_F$ distribution is larger at intermediate $x_F$, especially when the
7-particle configurations are included.  At larger $x_F$, the baryon
distributions always lead over charm mesons produced in the same configuration,
causing the $D^-/\Xi_c^0$ asymmetry to approach $-1$ as $x_F \rightarrow 1$.
Similar results should be expected from the models of Refs.~\cite{MB,SPetal}.

\begin{figure}[htb]
\setlength{\epsfxsize=0.95\textwidth}
\setlength{\epsfysize=0.5\textheight}
\centerline{\epsffile{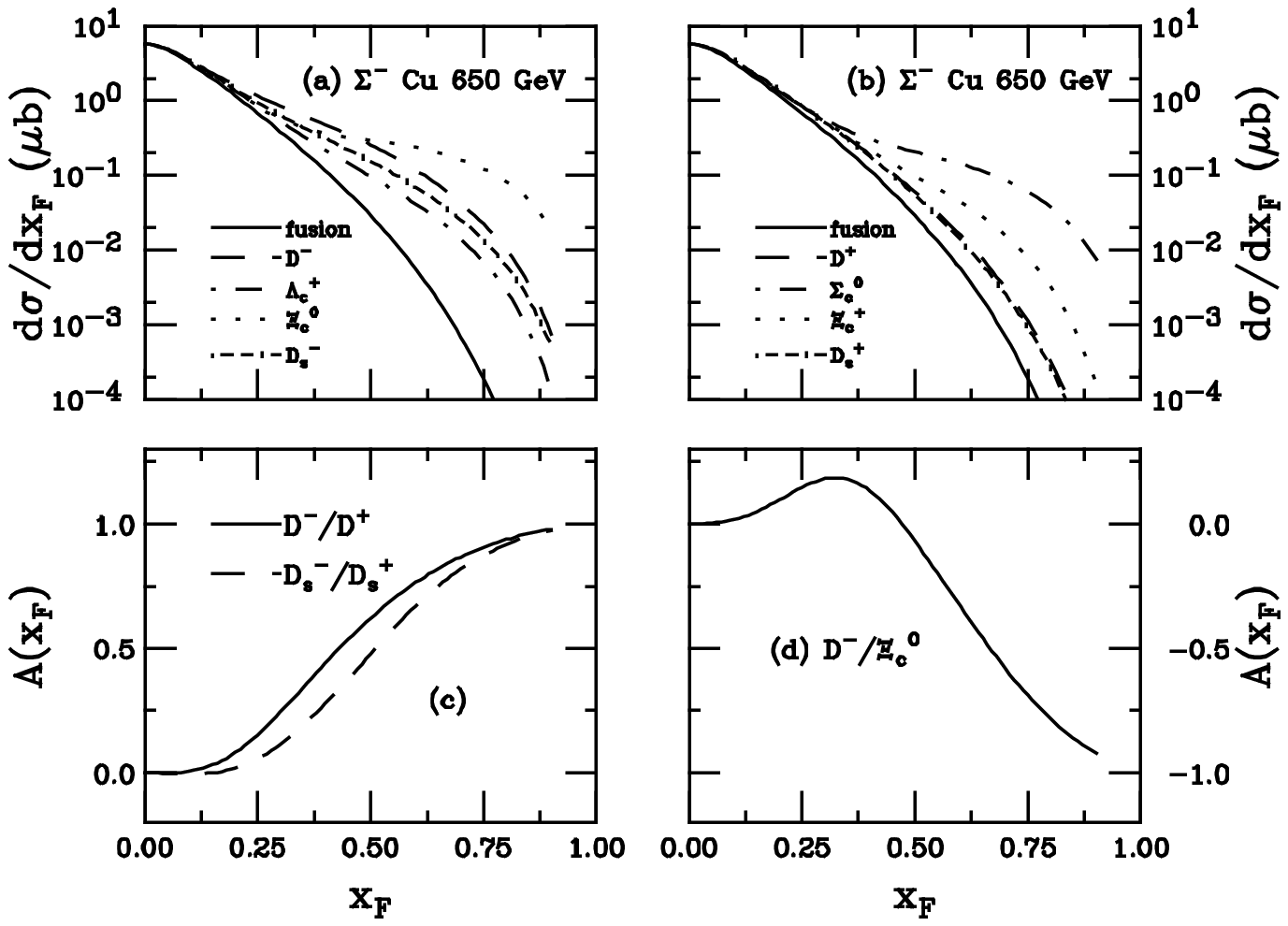}}
\caption[]{ Predictions for charm hadron production are given in our full model
for $\Sigma^-$Cu interactions at 650 GeV.  
The individual $x_F$ distributions are given in
(a) and (b).  All cross sections are compared to the leading twist fusion
calculation in the solid curve.  In (a) the hadron distributions are $D^-$
(dashed), $\Lambda_c^+$ (dot-dashed), $\Xi_c^0$ (dotted) and $D_s^-$ (dot
dashed dashed). In (b) the hadron distributions are $D^+$
(dashed), $\Sigma_c^0$ (dot-dashed), $\Xi_c^+$ (dotted) and $D_s^+$ (dot
dashed dashed).  Predictions of the asymmetries are given in (c) for
$D^-/D^+$ (solid) and $D_s^-/D_s^+$ (dashed) while the  prediction for the 
$D^-/\Xi_c^0$ asymmetry is given in (d).}
\label{sig650}
\end{figure}

We now turn to predictions of charm hadron production at SELEX with 650 GeV
beams of $\Sigma^-$ and $\pi^-$ \cite{SELEX}.  First we give the charm hadron
$x_F$ distributions for $\Sigma^-$Cu interactions and the relevant asymmetries
in Fig.~\ref{sig650}.  Since the leading-twist fusion cross section grows
faster than $\sigma_{\rm ic}^H$, the average $x_F$ of the particles studied
decreases $\approx 30$\% from 330 GeV to 650 GeV.  A smaller decrease is found
from the PYTHIA model, showing the relative strength of the string
fragmentation mechanism, as can be seen in Table~\ref{avexffull}.  The
$\Xi_c^0$ is clearly the hardest distribution, followed by the $\Sigma_c^0$.
The $\Xi_c^0$ leads the $\Sigma_c^0$ because the more massive valence $s$ quark
carries more of the $\Sigma^-$ velocity than the $d$ valence quarks.  The
$\Xi_c^+$ leads the $\Lambda_c^+$ in the 7-particle $u\overline u$ state for
the same reason.  The $D^-$ and $D_s^-$, also produced from the 5-particle
state have the hardest meson distributions but lag the baryons.  The $D^+$ and
$D_s^+$ have the softest distributions with the $D^+$ slightly harder because
the quarks in the $d \overline d$ configuration get slightly more velocity than
the $s \overline s$ configuration with the more massive strange quarks.  The
asymmetries, which should be compared to the dashed curves in
Fig.~\ref{asym330}, are somewhat reduced at higher energies, again due to the
larger leading-twist cross section.

\begin{figure}[htb]
\setlength{\epsfxsize=0.95\textwidth}
\setlength{\epsfysize=0.5\textheight}
\centerline{\epsffile{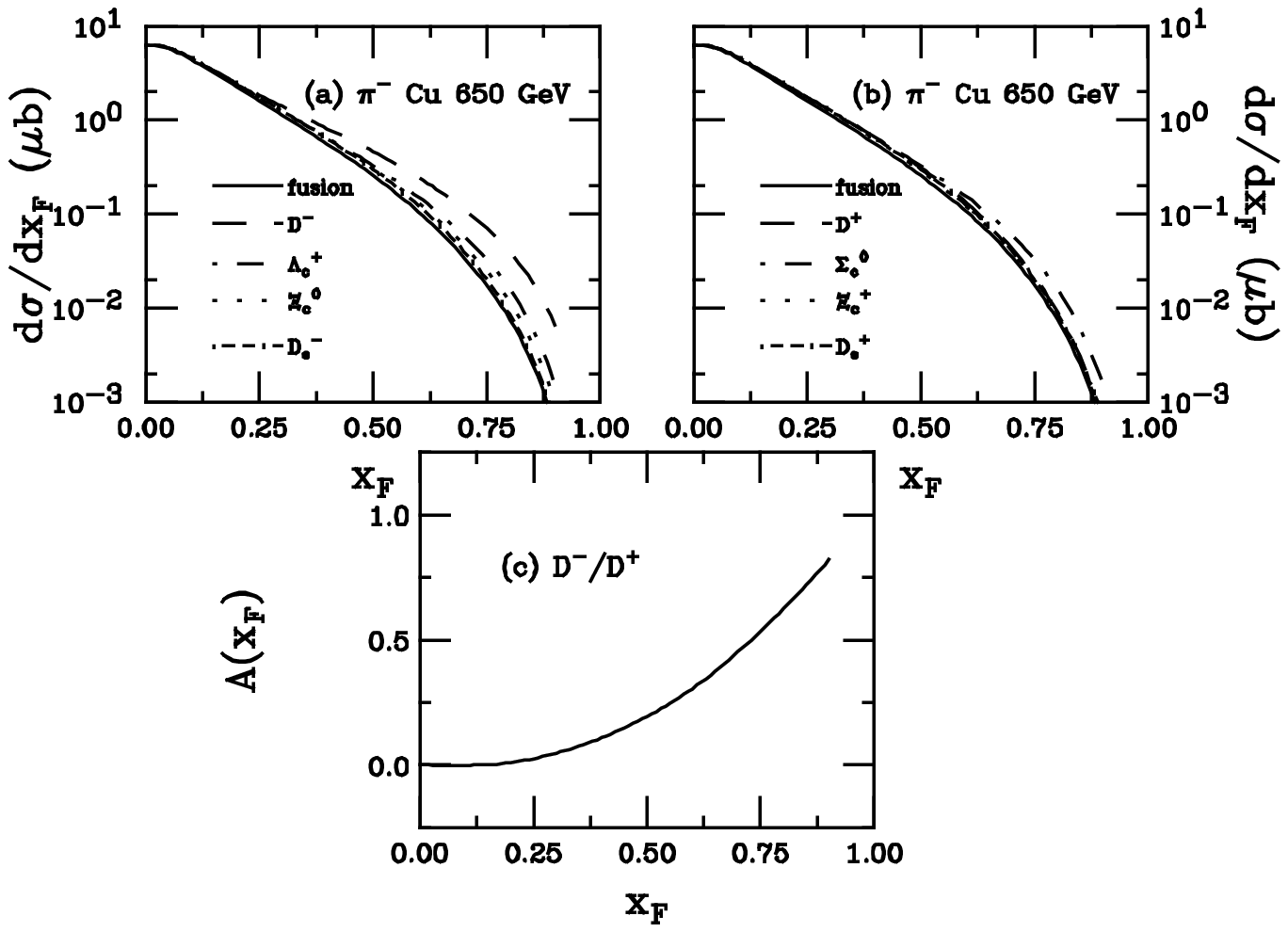}}
\caption[]{ Predictions for charm hadron production are given in our full model
for 650 GeV $\pi^-$Cu interactions.  
The individual $x_F$ distributions are given in
(a) and (b).  All cross sections are compared to the leading twist fusion
calculation in the solid curve.  In (a) the hadron distributions are $D^-$
(dashed), $\Lambda_c^+$ (dot-dashed), $\Xi_c^0$ (dotted) and $D_s^-$ (dot
dashed dashed). In (b) the hadron distributions are $D^+$
(dashed), $\Sigma_c^0$ (dot-dashed), $\Xi_c^+$ (dotted) and $D_s^+$ (dot
dashed dashed).  A prediction of the $D^-/D^+$ asymmetry is given in (c).}
\label{pi650}
\end{figure}

Since SELEX will also measure charm hadroproduction with a $\pi^-$ beam at the
same energy, these predictions are shown in Fig.~\ref{pi650}.  Because only the
$D^-$ is produced from the minimal Fock state configuration, it shows the
hardest $x_F$ distribution in Fig.~\ref{pi650}(a).  Note that since $c
\overline c$ production by leading-twist fusion alone is already significantly
harder than the equivalent production by baryon projectiles, the distributions
produced by coalescence from 6-particle configurations are not substantially
enhanced over the fusion cross section, even at large $x_F$.  The intrinsic
charm cross section is proportional to $\sigma_{\pi N}^{\rm in}$ which
increases slowly compared to the leading-twist cross section, further
decreasing the predicted leading charm enhancement.  Additionally, we note that
charm baryons lead mesons produced by coalescence only in the 6-particle
configurations since the baryons take $\approx 50$\% of the pion momentum while
the mesons take less, as seen in Fig.~\ref{piic} and Table~\ref{icavexf}.  The
PYTHIA distributions in Fig.~\ref{pifus}, aside from the
leading $D^-$, are more central, also evident from Table~\ref{avexffull}.  In
Fig.~\ref{pi650}(c), only the $D^-/D^+$ asymmetry is shown because the model
predicts identical $D_s^-$ and $D_s^+$ meson and $\Lambda_c$ and $\overline
\Lambda_c^+$ baryon distributions, see the appendix, hence
no asymmetry.  We note that the asymmetry is reduced compared to calculations
at lower energy \cite{VB}.

\begin{figure}[htb]
\setlength{\epsfxsize=0.95\textwidth}
\setlength{\epsfysize=0.5\textheight}
\centerline{\epsffile{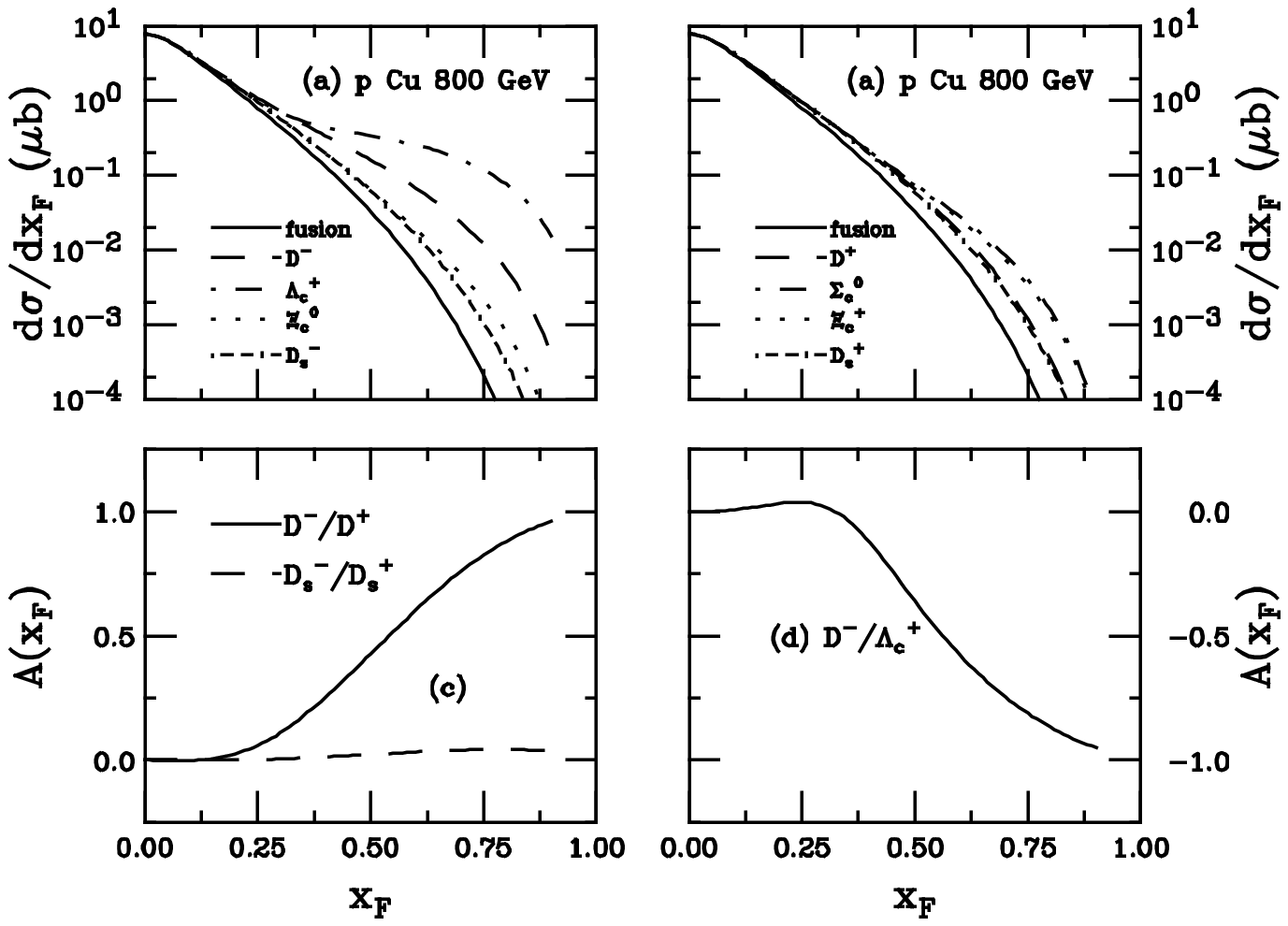}}
\caption[]{ Predictions for charm hadron production are given in our full model
for $p$Cu interactions at 800 GeV.  
The individual $x_F$ distributions are given in
(a) and (b).  All cross sections are compared to the leading twist fusion
calculation in the solid curve.  In (a) the hadron distributions are $D^-$
(dashed), $\Lambda_c^+$ (dot-dashed), $\Xi_c^0$ (dotted) and $D_s^-$ (dot
dashed dashed). In (b) the hadron distributions are $D^+$
(dashed), $\Sigma_c^0$ (dot-dashed), $\Xi_c^+$ (dotted) and $D_s^+$ (dot
dashed dashed).  Predictions of the asymmetries are given in (c) for
$D^-/D^+$ (solid) and $D_s^-/D_s^+$ (dashed) while the  prediction for the
$D^-/\Lambda_c^+$ asymmetry is given in (d).}
\label{pro800}
\end{figure}

The primary proton beam for fixed-target experiments at Fermilab is 800 GeV so
for completeness, we also give predictions for a possible $pA$ measurement at
this energy in Fig.~\ref{pro800}.  In this case, the $\Lambda_c^+$ has the
hardest $x_F$ distribution followed by the $D^-$, both of which are produced by
coalescence from the 5-particle Fock state.  Again, the $\Sigma_c^0$ and $D^+$
are somewhat harder than the $\Xi_c^+$ and $D_s^+$ distributions respectively
due to the relative partitioning of the parton velocity in the 7-particle $u
\overline u$ and $d \overline d$ configurations compared to the 7-particle
$s \overline s$ state.  The model predicts a strong $D^-/D^+$ asymmetry as well
as a $D^-/\Lambda_c^+$ asymmetry, comparable to the $D^-/\Xi_c^0$ asymmetry
predicted for the $\Sigma^-A$ interactions.  On the other hand, the
$D_s^-/D_s^+$ asymmetry is quite weak.  Such measurements with a proton beam
would provide a useful complement to a high statistics $\Sigma^-$ measurement.
A comprehensive understanding of data with proton projectiles has 
suffered in the past from a lack of statistics
and high precision proton data, compared to that from $\Sigma^-$ and $\pi^-$ 
projectiles, could eliminate certain models.

\section{Summary and Conclusions}

We have refined the intrinsic charm model of Refs.\
\cite{VB,VBlam,VBH2}, including both the minimal Fock state and all the
configurations with an additional $q \overline q$ pair.  We have applied a
simple counting scheme to determine the relative contribution of each state to
the final charm hadron distribution.  The model compares rather favorably to
the $x_F$ distributions measured by WA89 \cite{wa89} and produces reasonable
agreement with their measured $D^-/D^+$ asymmetry while falling short of the
$D_s^-/D_s^+$ and $\Lambda_c/\overline \Lambda_c$ data \cite{wa89asym}
at intermediate $x_F$. 

Further, we have made predictions for charm hadron production at the energy of
SELEX for both $\Sigma^-$ and $\pi^-$ projectiles.  Predictions for production
by an 800 GeV proton beam are also given.  High statistics data on charm
production from a combination of these projectiles could eliminate certain
classes of models and perhaps distinguish between coalescence in the initial
state, as in the intrinsic charm model, and in the final state, as in models
such as PYTHIA \cite{PYT}.  The simple counting scheme employed here could be
replaced with relative rates from data.  However, the shapes of the
distributions would not change significantly in our model.  Therefore a
collection of charm production data could define the role of intrinsic charm in
future experiments.

Acknowledgments: We thank S.J. Brodsky, E. Ramberg, H.-W. Siebert and 
T. Sj\"{o}strand for discussions.

\newpage
\setcounter{equation}{0}
\renewcommand{\theequation}{A-\arabic{equation}}
\begin{center}
{\bf Appendix}
\end{center}
\vspace{0.2in}

Here we give the probability distributions for $D^-$, $D^+$, $D_s^-$ and
$D_s^+$ mesons and $\Lambda_c^+$, $\Sigma_c^0$, $\Xi_c^0$ and $\Xi_c^+$
baryons for production by the minimal and first three higher Fock state
configurations from $\Sigma^-$, proton and $\pi^-$ projectiles.  The
probability distributions for each final state are given in
Figs.~\ref{sigic}-\ref{piic}.  We note that the predictions for $\Lambda_c^+$
and $\Sigma_c^+$ are identical in all cases because their quark content is the
same. Recall that $P_{\rm ic}^5 = 0.31$\%, $P_{\rm icu}^7 = P_{\rm icd}^7 =
70.4$\% $P_{\rm ic}^5$ and  $P_{\rm ics}^7 = 28.5$\% $P_{\rm ic}^5$.

We begin with the $\Sigma^-$. In the
$|ddsc \overline c \rangle$ configuration, there are four
final-state hadrons with a valence $c$ quark (2$\Xi_c^0$, $\Sigma_c^0$ and
$J/\psi$) and also four final-state hadrons with a valence $\overline c$ quark
(2$D^-$, $D_s^-$ and $J/\psi$). The $|ddsc \overline c q \overline q \rangle$ configurations where
$q \overline q = u \overline u$, $d \overline d$ and $s \overline s$ allow
coalescence production of eight possible final-state hadrons with a 
valence $c$ and five possible
final-state hadrons with a valence $\overline c$.  When $q \overline q = u
\overline u$, the possible hadrons produced by coalescence are 2$\Xi_c^+$,
$\Xi_c^0$, 2$\Lambda_c^+$, $\Sigma_c^0$, $D^0$ and $J/\psi$ with a valence $c$
and 2$D^-$, $D_s^-$, $\overline D^0$ and $J/\psi$ with a valence $\overline c$.
The $d \overline d$ configuration
allows coalescence production of the following hadrons
with a valence $c$ quark, 3$\Sigma_c^0$, 3$\Xi_c^0$, $D^+$
and $J/\psi$, and, with a valence $\overline c$, 3$D^-$, $D_s^-$ and $J/\psi$.
The $s \overline s$ configuration yields
4$\Xi_c^0$, $\Sigma_c^0$, $\Omega_c^0(ssc)$, $D_s^+$ and $J/\psi$
while the final-state valence $\overline c$
quarks hadrons are 2$D^-$, 2$D_s^-$ and $J/\psi$.  We have: 
\be \frac{dP_{D^-}}{dx_F} & = & \frac{1}{2} \left( \frac{1}{10} \frac{dP_{\rm
ic}^{5F}}{dx_F} + \frac{1}{2} \frac{dP_{\rm ic}^{5C}}{dx_F} \right) + 
 \frac{1}{2} \left( \frac{1}{10} \frac{dP_{\rm
icu}^{7F}}{dx_F} + \frac{2}{5} \frac{dP_{\rm icu}^{7C}}{dx_F} \right) \nonumber
\\ &   & + \, \frac{1}{2} \left( \frac{1}{10} \frac{dP_{\rm
icd}^{7F}}{dx_F} + \frac{3}{5} \frac{dP_{\rm icd}^{7C}}{dx_F} \right) +
 \frac{1}{2} \left( \frac{1}{10} \frac{dP_{\rm
ics}^{7F}}{dx_F} + \frac{2}{5} \frac{dP_{\rm ics}^{7C}}{dx_F} \right) 
\label{sigprobdm} \\
 \frac{dP_{D^+}}{dx_F} & = & \frac{1}{10} \frac{dP_{\rm ic}^{5F}}{dx_F} +
 \frac{1}{10} \frac{dP_{\rm icu}^{7F}}{dx_F} + \frac{1}{2} \left( \frac{1}{10}
\frac{dP_{\rm icd}^{7F}}{dx_F} + \frac{1}{8} \frac{dP_{\rm icd}^{7C}}{dx_F} 
\right) + \frac{1}{10} \frac{dP_{\rm ics}^{7F}}{dx_F} \, \, 
\label{sigprobdp} \\
 \frac{dP_{\Lambda_c^+}}{dx_F} & = & \frac{1}{10} \frac{dP_{\rm 
ic}^{5F}}{dx_F} + \frac{1}{2} \left( \frac{1}{10}
\frac{dP_{\rm icu}^{7F}}{dx_F} + \frac{1}{4} \frac{dP_{\rm icu}^{7C}}{dx_F} 
\right) + \frac{1}{10} \frac{dP_{\rm icd}^{7F}}{dx_F} + \frac{1}{10} 
\frac{dP_{\rm ics}^{7F}}{dx_F} \, \, 
\label{sigproblc} \\
 \frac{dP_{\Sigma_c^0}}{dx_F} & = & \frac{1}{2} \left( \frac{1}{10} 
\frac{dP_{\rm
ic}^{5F}}{dx_F} + \frac{1}{4} \frac{dP_{\rm ic}^{5C}}{dx_F} \right) + 
 \frac{1}{2} \left( \frac{1}{10} \frac{dP_{\rm
icu}^{7F}}{dx_F} + \frac{1}{8} \frac{dP_{\rm icu}^{7C}}{dx_F} \right) \nonumber
\\ &   & + \, \frac{1}{2} \left( \frac{1}{10} \frac{dP_{\rm
icd}^{7F}}{dx_F} + \frac{3}{8} \frac{dP_{\rm icd}^{7C}}{dx_F} \right) +
 \frac{1}{2} \left( \frac{1}{10} \frac{dP_{\rm
ics}^{7F}}{dx_F} + \frac{1}{8} \frac{dP_{\rm ics}^{7C}}{dx_F} \right) 
\label{sigprobs0} \\
 \frac{dP_{D_s^-}}{dx_F} & = & \frac{1}{2} \left( \frac{1}{10} \frac{dP_{\rm
ic}^{5F}}{dx_F} + \frac{1}{4} \frac{dP_{\rm ic}^{5C}}{dx_F} \right) + 
 \frac{1}{2} \left( \frac{1}{10} \frac{dP_{\rm
icu}^{7F}}{dx_F} + \frac{1}{5} \frac{dP_{\rm icu}^{7C}}{dx_F} \right) \nonumber
\\ &   & + \, \frac{1}{2} \left( \frac{1}{10} \frac{dP_{\rm
icd}^{7F}}{dx_F} + \frac{1}{5} \frac{dP_{\rm icd}^{7C}}{dx_F} \right) +
 \frac{1}{2} \left( \frac{1}{10} \frac{dP_{\rm
ics}^{7F}}{dx_F} + \frac{2}{5} \frac{dP_{\rm ics}^{7C}}{dx_F} \right) 
\label{sigprobdsm} \\
 \frac{dP_{D_s^+}}{dx_F} & = & \frac{1}{10} \frac{dP_{\rm ic}^{5F}}{dx_F} +
 \frac{1}{10} \frac{dP_{\rm icu}^{7F}}{dx_F} + \frac{1}{10} \frac{dP_{\rm 
icd}^{7F}}{dx_F} + \frac{1}{2} \left( \frac{1}{10}
\frac{dP_{\rm ics}^{7F}}{dx_F} + \frac{1}{8} \frac{dP_{\rm ics}^{7C}}{dx_F} 
\right) \, \, 
\label{sigprobdsp} \\
 \frac{dP_{\Xi_c^0}}{dx_F} & = & \frac{1}{2} \left( \frac{1}{10} 
\frac{dP_{\rm
ic}^{5F}}{dx_F} + \frac{1}{2} \frac{dP_{\rm ic}^{5C}}{dx_F} \right) + 
 \frac{1}{2} \left( \frac{1}{10} \frac{dP_{\rm
icu}^{7F}}{dx_F} + \frac{1}{8} \frac{dP_{\rm icu}^{7C}}{dx_F} \right) \nonumber
\\ &   & + \, \frac{1}{2} \left( \frac{1}{10} \frac{dP_{\rm
icd}^{7F}}{dx_F} + \frac{3}{8} \frac{dP_{\rm icd}^{7C}}{dx_F} \right) +
 \frac{1}{2} \left( \frac{1}{10} \frac{dP_{\rm
ics}^{7F}}{dx_F} + \frac{1}{2} \frac{dP_{\rm ics}^{7C}}{dx_F} \right) 
\label{sigprobx0} \\
 \frac{dP_{\Xi_c^+}}{dx_F} & = & \frac{1}{10} \frac{dP_{\rm 
ic}^{5F}}{dx_F} + \frac{1}{2} \left( \frac{1}{10}
\frac{dP_{\rm icu}^{7F}}{dx_F} + \frac{1}{4} \frac{dP_{\rm icu}^{7C}}{dx_F} 
\right) + \frac{1}{10} \frac{dP_{\rm icd}^{7F}}{dx_F} + \frac{1}{10} 
\frac{dP_{\rm ics}^{7F}}{dx_F} \, \, . 
\label{sigprobxp}
\ee

Fewer charm hadrons are produced by coalescence from the
five-quark configuration
of the proton since it has no valence strange quark.  In the
$|uudc \overline c \rangle$ configuration, there are four
final-state hadrons with a valence $c$ quark (2$\Lambda_c^+$,
$\Sigma_c^{++}(uuc)$ and
$J/\psi$) and also four final-state hadrons with a valence $\overline c$ quark
(2$\overline D^0$, $D^-$ and $J/\psi$). The $|uudc \overline c q \overline q 
\rangle$ configurations allow
coalescence production of eight possible final-state hadrons with a 
valence $c$ and five possible 
final-state hadrons with a valence $\overline c$.  When $q \overline q = u
\overline u$, the possible hadrons produced by coalescence are: 
3$\Lambda_c^+$, 3$\Sigma_c^{++}$, $D^0$ and $J/\psi$ with a valence $c$
and $D^-$, 3$\overline D^0$ and $J/\psi$ with a valence $\overline c$.
The $d \overline d$ configuration
allows coalescence production of the following hadrons
with a valence $c$ quark, 4$\Lambda_c^+$, $\Sigma_c^0$, $\Sigma_c^{++}$, $D^+$
and $J/\psi$, and, with a valence $\overline c$, 2$D^-$, 2$\overline D^0$ 
and $J/\psi$.  The $s \overline s$ configuration yields
2$\Xi_c^+$, $\Xi_c^0$, 2$\Lambda_c^+$, $\Sigma_c^{++}$, $D_s^+$ and $J/\psi$
while the final-state valence $\overline c$
quarks hadrons are 2$D^-$, $\overline D^0$, $D_s^-$ and $J/\psi$.  Then:
\be \frac{dP_{D^-}}{dx_F} & = & \frac{1}{2} \left( \frac{1}{10} \frac{dP_{\rm
ic}^{5F}}{dx_F} + \frac{1}{4} \frac{dP_{\rm ic}^{5C}}{dx_F} \right) + 
 \frac{1}{2} \left( \frac{1}{10} \frac{dP_{\rm
icu}^{7F}}{dx_F} + \frac{1}{5} \frac{dP_{\rm icu}^{7C}}{dx_F} \right) \nonumber
\\ &   & + \, \frac{1}{2} \left( \frac{1}{10} \frac{dP_{\rm
icd}^{7F}}{dx_F} + \frac{2}{5} \frac{dP_{\rm icd}^{7C}}{dx_F} \right) +
 \frac{1}{2} \left( \frac{1}{10} \frac{dP_{\rm
ics}^{7F}}{dx_F} + \frac{1}{5} \frac{dP_{\rm ics}^{7C}}{dx_F} \right) \\
 \frac{dP_{D^+}}{dx_F} & = & \frac{1}{10} \frac{dP_{\rm ic}^{5F}}{dx_F} +
 \frac{1}{10} \frac{dP_{\rm icu}^{7F}}{dx_F} + \frac{1}{2} \left( \frac{1}{10}
\frac{dP_{\rm icd}^{7F}}{dx_F} + \frac{1}{8} \frac{dP_{\rm icd}^{7C}}{dx_F} 
\right) + \frac{1}{10} \frac{dP_{\rm ics}^{7F}}{dx_F} \, \, \\
 \frac{dP_{\Lambda_c^+}}{dx_F} & = & \frac{1}{2} \left( \frac{1}{10} 
\frac{dP_{\rm
ic}^{5F}}{dx_F} + \frac{1}{2} \frac{dP_{\rm ic}^{5C}}{dx_F} \right) + 
 \frac{1}{2} \left( \frac{1}{10} \frac{dP_{\rm
icu}^{7F}}{dx_F} + \frac{3}{8} \frac{dP_{\rm icu}^{7C}}{dx_F} \right) \nonumber
\\ &   & + \, \frac{1}{2} \left( \frac{1}{10} \frac{dP_{\rm
icd}^{7F}}{dx_F} + \frac{1}{2} \frac{dP_{\rm icd}^{7C}}{dx_F} \right) +
 \frac{1}{2} \left( \frac{1}{10} \frac{dP_{\rm
ics}^{7F}}{dx_F} + \frac{1}{4} \frac{dP_{\rm ics}^{7C}}{dx_F} \right) \\
 \frac{dP_{\Sigma_c^0}}{dx_F} & = & \frac{1}{10} \frac{dP_{\rm 
ic}^{5F}}{dx_F} + \frac{1}{10} \frac{dP_{\rm icu}^{7F}}{dx_F} + 
\frac{1}{2} \left( \frac{1}{10}
\frac{dP_{\rm icd}^{7F}}{dx_F} + \frac{1}{8} \frac{dP_{\rm icd}^{7C}}{dx_F} 
\right) + \frac{1}{10} 
\frac{dP_{\rm ics}^{7F}}{dx_F} \, \, \\
 \frac{dP_{D_s^-}}{dx_F} & = & \frac{1}{10} \frac{dP_{\rm ic}^{5F}}{dx_F} +
 \frac{1}{10} \frac{dP_{\rm icu}^{7F}}{dx_F} + \frac{1}{10} \frac{dP_{\rm 
icd}^{7F}}{dx_F} + \frac{1}{2} \left( \frac{1}{10}
\frac{dP_{\rm ics}^{7F}}{dx_F} + \frac{1}{5} \frac{dP_{\rm ics}^{7C}}{dx_F} 
\right) \, \, \\
 \frac{dP_{D_s^+}}{dx_F} & = & \frac{1}{10} \frac{dP_{\rm ic}^{5F}}{dx_F} +
 \frac{1}{10} \frac{dP_{\rm icu}^{7F}}{dx_F} + \frac{1}{10} \frac{dP_{\rm 
icd}^{7F}}{dx_F} + \frac{1}{2} \left( \frac{1}{10}
\frac{dP_{\rm ics}^{7F}}{dx_F} + \frac{1}{8} \frac{dP_{\rm ics}^{7C}}{dx_F} 
\right) \, \, \\
 \frac{dP_{\Xi_c^0}}{dx_F} & = & \frac{1}{10} \frac{dP_{\rm ic}^{5F}}{dx_F} +
 \frac{1}{10} \frac{dP_{\rm icu}^{7F}}{dx_F} + \frac{1}{10} \frac{dP_{\rm 
icd}^{7F}}{dx_F} + \frac{1}{2} \left( \frac{1}{10}
\frac{dP_{\rm ics}^{7F}}{dx_F} + \frac{1}{8} \frac{dP_{\rm ics}^{7C}}{dx_F} 
\right) \, \, \\
 \frac{dP_{\Xi_c^+}}{dx_F} & = & \frac{1}{10} \frac{dP_{\rm ic}^{5F}}{dx_F} +
 \frac{1}{10} \frac{dP_{\rm icu}^{7F}}{dx_F} + \frac{1}{10} \frac{dP_{\rm 
icd}^{7F}}{dx_F} + \frac{1}{2} \left( \frac{1}{10}
\frac{dP_{\rm ics}^{7F}}{dx_F} + \frac{1}{4} \frac{dP_{\rm ics}^{7C}}{dx_F} 
\right) \, \, . 
\label{proprob} 
\ee

Charm and anticharm hadron production is more symmetric from the $\pi^-$
because the projectile 
contains a valence antiquark of its own.  In the minimal
$|\overline u dc \overline c \rangle$ configuration, there are two
possible final-state hadrons with a valence $c$ quark ($D^0$
and $J/\psi$) and also two possible
final-state hadrons with a valence $\overline c$ quark
($D^-$ and $J/\psi$). The $|\overline u dc \overline c q \overline q 
\rangle$ configurations allow
coalescence production of four possible 
final-state hadrons with a valence $c$ and likewise four possible 
final-state hadrons with a valence $\overline c$.  When $q \overline q = u
\overline u$, the possible hadrons produced by coalescence are: 
$\Lambda_c^+$, 2$D^0$ and $J/\psi$ with a valence $c$
and $D^-$, $\overline D^0$, $\overline \Sigma_c^{++}$ and $J/\psi$ with a 
valence $\overline c$.  The $d \overline d$ configuration
allows coalescence production of the following hadrons
with a valence $c$ quark, $\Sigma_c^0$, $D^0$, $D^+$
and $J/\psi$, and, with a valence $\overline c$, $D^-$, $\overline D^0$,
$\overline \Lambda_c^+$ 
and $J/\psi$.  The $s \overline s$ configuration yields
$\Xi_c^0$, $D_s^+$, $D^0$ and $J/\psi$ with a valence $c$
while the possible final-state valence $\overline c$
quarks hadrons are $D^-$, $D_s^-$, $\overline \Xi_c^+$ and $J/\psi$. In this
case: 
\be \frac{dP_{D^-}}{dx_F} & = & \frac{1}{2} \left( \frac{1}{10} \frac{dP_{\rm
ic}^{4F}}{dx_F} + \frac{1}{2} \frac{dP_{\rm ic}^{4C}}{dx_F} \right) + 
 \frac{1}{2} \left( \frac{1}{10} \frac{dP_{\rm
icu}^{6F}}{dx_F} + \frac{1}{4} \frac{dP_{\rm icu}^{6C}}{dx_F} \right) \nonumber
\\ &   & + \, \frac{1}{2} \left( \frac{1}{10} \frac{dP_{\rm
icd}^{6F}}{dx_F} + \frac{1}{2} \frac{dP_{\rm icd}^{6C}}{dx_F} \right) +
 \frac{1}{2} \left( \frac{1}{10} \frac{dP_{\rm
ics}^{6F}}{dx_F} + \frac{1}{4} \frac{dP_{\rm ics}^{6C}}{dx_F} \right) \\
 \frac{dP_{D^+}}{dx_F} & = & \frac{1}{10} \frac{dP_{\rm ic}^{4F}}{dx_F} +
 \frac{1}{10} \frac{dP_{\rm icu}^{6F}}{dx_F} + \frac{1}{2} \left( \frac{1}{10}
\frac{dP_{\rm icd}^{6F}}{dx_F} + \frac{1}{4} \frac{dP_{\rm icd}^{6C}}{dx_F} 
\right) + \frac{1}{10} \frac{dP_{\rm ics}^{6F}}{dx_F} \, \, \\
 \frac{dP_{\Lambda_c^+}}{dx_F} & = & \frac{1}{10} \frac{dP_{\rm 
ic}^{4F}}{dx_F} + \frac{1}{2} \left( \frac{1}{10}
\frac{dP_{\rm icu}^{6F}}{dx_F} + \frac{1}{4} \frac{dP_{\rm icu}^{6C}}{dx_F} 
\right) + \frac{1}{10} \frac{dP_{\rm icd}^{6F}}{dx_F} + \frac{1}{10} 
\frac{dP_{\rm ics}^{6F}}{dx_F} \, \, \\
 \frac{dP_{\Sigma_c^0}}{dx_F} & = & \frac{1}{10} \frac{dP_{\rm 
ic}^{4F}}{dx_F} + \frac{1}{10} \frac{dP_{\rm icu}^{6F}}{dx_F} + 
\frac{1}{2} \left( \frac{1}{10}
\frac{dP_{\rm icd}^{6F}}{dx_F} + \frac{1}{4} \frac{dP_{\rm icd}^{6C}}{dx_F} 
\right) + \frac{1}{10} 
\frac{dP_{\rm ics}^{6F}}{dx_F} \, \, \\
 \frac{dP_{D_s^-}}{dx_F} & = & \frac{1}{10} \frac{dP_{\rm ic}^{4F}}{dx_F} +
 \frac{1}{10} \frac{dP_{\rm icu}^{6F}}{dx_F} + \frac{1}{10} \frac{dP_{\rm 
icd}^{6F}}{dx_F} + \frac{1}{2} \left( \frac{1}{10}
\frac{dP_{\rm ics}^{6F}}{dx_F} + \frac{1}{4} \frac{dP_{\rm ics}^{6C}}{dx_F} 
\right) \, \, \\
 \frac{dP_{D_s^+}}{dx_F} & = & \frac{1}{10} \frac{dP_{\rm ic}^{4F}}{dx_F} +
 \frac{1}{10} \frac{dP_{\rm icu}^{6F}}{dx_F} + \frac{1}{10} \frac{dP_{\rm 
icd}^{6F}}{dx_F} + \frac{1}{2} \left( \frac{1}{10}
\frac{dP_{\rm ics}^{6F}}{dx_F} + \frac{1}{4} \frac{dP_{\rm ics}^{6C}}{dx_F} 
\right) \, \, \\
 \frac{dP_{\Xi_c^0}}{dx_F} & = & \frac{1}{10} \frac{dP_{\rm ic}^{4F}}{dx_F} +
 \frac{1}{10} \frac{dP_{\rm icu}^{6F}}{dx_F} + \frac{1}{10} \frac{dP_{\rm 
icd}^{6F}}{dx_F} + \frac{1}{2} \left( \frac{1}{10}
\frac{dP_{\rm ics}^{6F}}{dx_F} + \frac{1}{4} \frac{dP_{\rm ics}^{6C}}{dx_F} 
\right) \, \, \\
 \frac{dP_{\Xi_c^+}}{dx_F} & = & \frac{1}{10} \frac{dP_{\rm ic}^{4F}}{dx_F} +
 \frac{1}{10} \frac{dP_{\rm icu}^{6F}}{dx_F} + \frac{1}{10} \frac{dP_{\rm 
icd}^{6F}}{dx_F} + \frac{1}{10} \frac{dP_{\rm ics}^{6F}}{dx_F} \, \, . 
\label{piprob}
\ee

\end{document}